\documentstyle[preprint,aps,epsf,floats,eqsecnum]{revtex}

\def\be{\begin{equation}}
\def\ee{\end{equation}}
\def\bea{\begin{eqnarray}}
\def\eea{\end{eqnarray}}
\def\lqcd{\Lambda_{\rm QCD}}
\def\mev{\,{\rm MeV}}
\def\gev{\,{\rm GeV}}
\def\case#1#2{\textstyle{{#1\over#2}}}
\def\lhqet{{\cal L}_{\rm HQET}}
\def\Dslash{\rlap{\,/}D}
\def\mbpole{m^{\rm pole}_b}
\def\mcpole{m^{\rm pole}_c}

\begin{document}

\preprint{\vbox{\hbox{JHU--TIPAC--96017}\hbox{hep-ph/9610363}
\hbox{October, 1996}}}

\title{The Heavy Quark Expansion of QCD}
\author{Adam F.~Falk}
\address{Department of Physics and Astronomy\\
The Johns Hopkins University\\
3400 North Charles Street, Baltimore, Maryland 21218}

\maketitle

\begin{abstract}
These lectures contain an elementary introduction to heavy quark symmetry and
the heavy quark expansion.  Applications such as the expansion of heavy meson
decay constants and the treatment of inclusive and exclusive semileptonic $B$
decays are included.  Heavy hadron production via nonperturbative
fragmentation processes is also discussed.
\end{abstract}

\vskip1in

{\it
\centerline{to appear in the}
\centerline{Proceedings of the XXIVth SLAC Summer Institute on Particle Physics}
\centerline{``The Strong Interaction, from Hadrons to Partons''}
\centerline{Stanford, California, August 19-30, 1996}}

\vfill\eject

\section{Heavy Quark Symmetry}

In these lectures I will introduce the ideas of heavy quark symmetry and the
heavy quark limit, which exploit the simplification of certain aspects of QCD for infinite quark mass, $m_Q\to\infty$.  We will see that while these ideas are extraordinarily simple from a physical point of view, they are of enormous practical utility
in the study of the phenomenology of bottom and charmed hadrons.  One reason
for this is the existence not just of an interesting new limit of QCD, but of a systematic expansion about this limit.  The technology of this
expansion is the Heavy Quark Effective Theory (HQET), which allows one to use heavy
quark symmetry to make accurate predictions of the properties and behavior of
heavy hadrons in which the theoretical errors are under control.  While the
emphasis in these lectures will be on the physical picture of heavy hadrons
which emerges in the heavy quark limit, it will be important to introduce
enough of the formalism of the HQET to reveal the structure of the heavy quark
expansion as a simultaneous expansion in powers of $\Lambda_{\rm QCD}/m_Q$ and
$\alpha_s(m_Q)$.  However, what I hope to leave you with above all is an
appreciation for the simplicity, elegance and coherence of the ideas which
underlie the technical results which will be presented.  The interested reader is also encouraged to consult a number of excellent reviews~\cite{reviews}, which typically cover in more detail the material in the first two sections of these lectures.

\subsection{Introduction}

Why is an understanding of QCD crucial to the study of the properties of the bottom and charmed quarks?  As an example, consider semileptonic $b$ decay,
$b\to c\,\ell\bar\nu$.  This process is mediated by a four-fermion operator,
\be\label{Obcdef}
  {\cal O}_{bc}={G_FV_{cb}\over\sqrt2}\bar c\gamma^\mu(1-\gamma^5)b\,
  \bar\nu\gamma_\mu(1-\gamma^5)\ell\,.
\ee
The weak matrix element is easy to calculate at the quark level,
\be
  {\cal A}_{\rm quark} = \langle c\,\ell\bar\nu|\,{\cal O}_{bc}|b\rangle=
  {G_FV_{cb}\over\sqrt2}\bar u_c(p_c)\gamma^\mu(1-\gamma^5)u_b(p_b)\,
  \bar u_\ell(p_\ell)\gamma_\mu(1-\gamma^5)v_\nu(p_\nu)\,.
\ee
However, ${\cal A}_{\rm quark}$ is only relevant at very short distances; at longer distances, QCD confinement implies that free $b$ and $c$ quarks are not asymptotic states of the theory.  Instead, nonperturbative QCD effects ``dress'' the quark level transition $b\to c\,\ell\bar\nu$ to a hadronic transition, such as 
\be
  B\to D\ell\bar\nu\quad{\rm or}\quad B\to D^*\ell\bar\nu
  \quad{\rm or}\quad \ldots\,
\ee
(In these lectures, we will use a convention in which a $B$ meson contains a $b$ quark, not a $\bar b$ antiquark.)  The hadronic matrix element 
${\cal A}_{\rm hadron}$ depends on nonperturbative QCD as well as on $G_FV_{cb}$, and is difficult to calculate from first principles.  To disentangle the weak interaction part of this complicated process requires us to develop some understanding of the strong interaction effects.

There are a variety of methods by which one can do this.  Perhaps the most popular, historically, has been use of various quark potential models~\cite{models}.  While these models are typically very predictive, they are based on uncontrolled assumptions and approximations, and it is virtually impossible to estimate the theoretical errors associated with their use.  This is a serious defect if one builds such a model into the experimental extraction of a weak coupling constant such as $V_{cb}$, because the uncontrolled theoretical errors then infect the experimental result.

Another approach, the one to be discussed here, is to exploit as much as possible the fact that the $b$ and $c$ quarks are {\it heavy}, by which we mean that $m_b,m_c\gg\lqcd$.  The scale $\lqcd$ is the typical energy at which QCD becomes nonperturbative, and is of the order of hundreds of MeV.  The physical quark masses are approximately $m_b\approx4.8\gev$ and $m_c\approx1.5\gev$.  The formalism which we will develop will not make as many predictions as do potential models.  However, the compensation will be that we will develop a {\it systematic\/} expansion in powers of $\lqcd/m_{b,c}$, within which we will be able to do concrete error analysis.  In particular, we will be able to  estimate the error associated with the fact that $m_c$ may not be very close to the asymptotic limit $m_c\gg\lqcd$.  Even where this error may be substantial, the fact that it is under control allows us to maintain predictive power in the theory.

\subsection{The heavy quark limit}

Consider a hadron $H_Q$ composed of a heavy quark $Q$ and ``light degrees of freedom'', consisting of light quarks, light antiquarks and gluons, in the limit $m_Q\to\infty$.  The Compton wavelength of the heavy quark scales as the inverse of the heavy quark mass, $\lambda_Q\sim1/m_Q$.  The light degrees of freedom, by contrast, are characterized by momenta of order $\lqcd$, corresponding to wavelengths $\lambda_\ell\sim1/\lqcd$.  Since $\lambda_\ell\gg\lambda_Q$, the light degrees of freedom cannot resolve features of the heavy quark other than its conserved gauge quantum numbers.  In particular, they cannot probe the actual {\it value\/} of $\lambda_Q$, that is, the value of $m_Q$.

We draw the same conclusion in momentum space.  The structure of the hadron $H_Q$ is determined by nonperturbative strong interactions.  The asymptotic freedom of QCD implies that when quarks and gluons exchange momenta $p$ much larger than $\lqcd$, the process is perturbative in the strong coupling constant $\alpha_s(p)$.  On the other hand, the typical momenta exchanged by the light degrees of freedom with each other and with the heavy quark are of order $\lqcd$, for which a perturbative expansion is of no use.  For these exchanges, however, $p<m_Q$, and the heavy quark $Q$ does not recoil, remaining at rest in the rest frame of the hadron.  In this limit, $Q$ acts as a static source of electric and chromoelectric gauge field.  The chromoelectric field, which holds $H_Q$ together, is nonperturbative in nature, but it is independent of $m_Q$.  The result is that the properties of the light degrees of freedom depend only on the presence of the static gauge field, independent of the flavor and mass of the heavy quark carrying the gauge charge.\footnote{Top quarks decay too quickly for a static chromoelectric field to be established around them, so the simplifications discussed here are not relevant to them.}

There is an immediate implication for the spectroscopy of heavy hadrons.  Since the interaction of the light degrees of freedom with the heavy quark is independent of $m_Q$, then so is the spectrum of their excitations.  It is these excitations which determine the spectrum of heavy hadrons $H_Q$.  Hence the {\it splittings\/} $\Delta_i\sim\lqcd$ between the various hadrons $H_Q^i$ are independent of $Q$ and, in the limit $m_Q\to\infty$, do not scale with $m_Q$.

\begin{figure}
\epsfxsize=8cm
\hfil\epsfbox{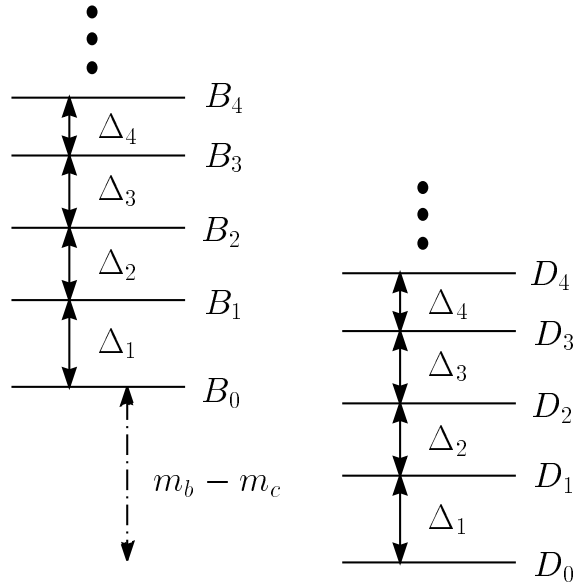}\hfill
\caption{Schematic spectra of the bottom and charmed mesons in the limit $m_b,m_c\gg\lqcd$.  The offset of the two spectra is not to scale; in reality, $m_b-m_c\gg\Delta_i\sim\lqcd$.}
\label{fig:spectra}
\end{figure}

For example, the spectra of bottom and charmed mesons are shown schematically in Fig.~\ref{fig:spectra}, in the limit $m_b,m_c\gg\lqcd$.  The light degrees of freedom are in exactly the same state in the mesons $B_i$ and $D_i$, for a given $i$.  The offset $B_i-D_i=m_b-m_c$ is just the difference between the heavy quark masses; in no way does the relationship between the spectra rely on an approximation $m_b\approx m_c$.

We can enrich this picture by recalling that the heavy quarks and light degrees of freedom also carry spin.  The heavy quark has spin quantum number $S_Q={1\over2}$, which leads to a chromomagnetic moment
\be
  \mu_Q={g\over2m_Q}\,.
\ee
Note that $\mu_Q\to0$ as $m_Q\to\infty$, and the interaction between the spin of the heavy quark and the light degrees of freedom is suppressed.  Hence the light degrees of freedom are insensitive to $S_Q$; their state is independent of whether $S_Q^z={1\over2}$ or $S_Q^z=-{1\over2}$.  Thus each of the energy levels in Fig.~\ref{fig:spectra} is actually doubled, one state for each possible value of  $S_Q^z$.

To summarize, what we see is that the light degrees of freedom are the same when combined with any of the following heavy quark states:
\be
  Q_1(\uparrow)\,,\quad Q_1(\downarrow);\quad 
  Q_2(\uparrow)\,,\quad Q_2(\downarrow);\quad\dots\quad
  Q_{N_h}(\uparrow)\,,\quad Q_{N_h}(\downarrow)\,,
\ee
where there are $N_h$ heavy quarks (in the real world, $N_h=2$).  The result is an $SU(2N_h)$ symmetry which applies to the light degrees of freedom~\cite{IW,VS,NW,PW,ES}.  A new symmetry means new nonperturbative relations between physical quantities.  It is these relations which we wish to understand and exploit.

The light degrees of freedom have total angular momentum $J_\ell$, which is integral for baryons and half-integral for mesons.  When combined with the 
heavy quark spin $S_Q={1\over2}$, we find physical hadron states with total angular momentum
\be
  J=\left| J_\ell\pm\case12 \right|\,.
\ee
If $S_\ell\ne0$, then these two states are degenerate.  For example, the lightest heavy mesons have $S_\ell=\case12$, leading to a doublet with $J=0$ and $J=1$.  Indeed, in the charm system we find that the states of lowest mass are the spin-0 $D$ and the spin-1 $D^*$; the corresponding bottom mesons are the $B$ and $B^*$.  

When effects of order $1/m_Q$ are included, the chromomagnetic interactions split the states of given $S_\ell$ but different $J$.  This ``hyperfine'' splitting is not calculable perturbatively, but it is proportional to the heavy quark magnetic moment $\mu_Q$.  This gives its scaling with $m_Q$:
\bea
  m_{D^*}-m_D &\sim& 1/m_c\nonumber\\
  m_{B^*}-m_B &\sim& 1/m_b\,.
\eea
From this fact we can construct a relation which is a nonperturbative prediction of heavy quark symmetry,
\be
  m_{B^*}^2-m_B^2=m_{D^*}^2-m_D^2\,.
\ee
Experimentally, $m_{B^*}^2-m_B^2=0.49\gev^2$ and $m_{D^*}^2-m_D^2=0.55\gev^2$, so this prediction works quite well.  Note that this relation involves not just the heavy quark symmetry, but the systematic inclusion of the leading symmetry violating effects.

So far, we have formulated heavy quark symmetry for hadrons in their rest frame.  Of course, we can easily boost to a frame in which the hadrons have arbitrary four-velocity $v^\mu=\gamma(1,\vec v)$.  For heavy quarks $Q_1$ and $Q_2$, the symmetry will then relate hadrons $H_1(v)$ and $H_2(v)$ with the same velocity but with different momenta.  This distinguishes heavy quark symmetry from ordinary symmetries of QCD, which relate states of the same momentum.  To remind ourselves of this distinction, henceforth we will label heavy hadrons explicitly by their velocity: $D(v)$, $D^*(v)$, $B(v)$, $B^*(v)$, and so on.

\subsection{Semileptonic decay of a heavy quark}

Now let us return to the semileptonic weak decay $b\to c\,\ell\bar\nu$, but now consider it in the heavy quark limit for the $b$ and $c$ quarks.  Suppose the decay occurs at time $t=0$.  For $t<0$, the $b$ quark is embedded in a hadron $H_b$; for $t>0$, the $c$ quark is dressed by light degrees of freedom to $H_c'$.  Let us consider the lightest hadrons, $H_b=B(v)$ and $H_c'=D(v')$.  Note that since the leptons carry away energy and momentum, in general $v\ne v'$.

What happens to the light degrees of freedom when the heavy quark decays?  For $t<0$, they see the chromoelectric field of a point source with velocity $v$.  At $t=0$, this point source recoils instantaneously\footnote{The weak decay occurs over a very short time $\delta t\sim1/M_W\ll1/\lqcd$.} to velocity $v'$; the color neutral leptons do not interact with the light hadronic degrees of freedom as they fly off.  The light quarks and gluons then must reassemble themselves about the recoiling color source.  This nonperturbative process will generally involve the production of an excited state or of additional particles; the light degrees of freedom can exchange energy with the heavy quark, so there is no kinematic restriction on the excitations (of energy $\sim\lqcd$) which can be formed.  There is also some chance that the light degrees of freedom will reassemble themselves back into a ground state $D$ meson.  The amplitude for this to happen is a function only of the inner product $w=v\cdot v'$ of the initial and final velocities of the color sources.  This amplitude, $\xi(w)$, is known as the Isgur-Wise function~\cite{IW}.

Clearly, the kinematic point $v=v'$, or $w=1$, is a special one.  In this corner of phase space, where the leptons are emitted back to back, there is no recoil of the source of color field at $t=0$.  As far as the light degrees of freedom are concerned, {\it nothing happens!\/}  Their state is unaffected by the decay of the heavy quark; they don't even notice it.  Hence the amplitude for them to remain in the ground state is exactly unity.  This is reflected in a nonperturbative normalization of the Isgur-Wise function at zero recoil~\cite{IW},
\be\label{norm}
  \xi(1)=1\,.
\ee
As we will see, this normalization condition is of enormous phenomenological use.  It will be extremely important to understand the corrections to this result for finite heavy quark masses $m_b$ and, especially, $m_c$.

The weak decay $b\to c$ is mediated by a left-handed current 
$\bar c\gamma^\mu(1-\gamma^5)b$.  Not only does this operator carry momentum, but it can change the orientation of the spin $S_Q$ of the heavy quark during the decay.  For a fixed light angular momentum $J_\ell$, the relative orientation of $S_Q$ determines whether the physical hadron in the final state is a $D$ or a $D^*$.  However, the light degrees of freedom are insensitive to $S_Q$, so the nonperturbative part of the transition is the same whether it is a $D$ or a $D^*$ which is produced.  Hence heavy quark symmetry implies relations between the hadronic matrix elements which describe the semileptonic decays $B\to D\ell\bar\nu$ and $B\to D^*\ell\bar\nu$.

It is conventional to parameterize these matrix elements by a set of scalar form factors.  These are defined separately for the vector and axial currents, as follows:
\bea\label{formfactors}
  \langle D(v')|\,\bar c\gamma^\mu b\,|B(v)\rangle &=&
  h_+(w)(v+v')^\mu+h_-(w)(v-v')^\mu\nonumber\\
  \langle D^*(v',\epsilon)|\,\bar c\gamma^\mu b\,|B(v)\rangle &=&
  h_V(w)i\varepsilon^{\mu\nu\alpha\beta}\epsilon_\nu^*v'_\alpha v_\beta
  \nonumber\\
  \langle D(v')|\,\bar c\gamma^\mu\gamma^5 b\,|B(v)\rangle &=& 0\\
  \langle D(v',\epsilon)|\,\bar c\gamma^\mu\gamma^5 b\,|B(v)\rangle &=&
  h_{A_1}(w)(w+1)\epsilon^{*\mu}-\epsilon^*\cdot v
  [h_{A_2}(w)v^\mu+h_{A_3}(w)v^{\prime\mu}]\,.\nonumber
\eea
The set of form factors $h_i(w)$ is the one appropriate to the heavy quark limit.  Other linear combinations are also found in the literature.  In any case, the form factors are independent nonperturbative functions of the recoil or equivalently, for fixed $m_b$ and $m_c$, of the momentum transfer.  However, in the heavy quark limit they correspond to a {\it single\/} transition of the light degrees of freedom, being distinguished from each other only by the relative orientation of the spin of the heavy quark.  Hence they may all be written in terms of the single function $\xi(w)$ which describes this nonperturbative transition.  As we will derive later, the result is a set of relations~\cite{IW},
\bea\label{ffrelations}
  &&h_+(w)=h_V(w)=h_{A_1}(w)=h_{A_3}(w)=\xi(w)\nonumber\\
  &&h_-(w)=h_{A_2}(w)=0\,,
\eea
which follow solely from the heavy quark symmetry.  Of course, all of the form factors which do not vanish inherit the normalization condition (\ref{norm}) at zero recoil.  This result is a powerful constraint on the structure of semileptonic decay in the heavy quark limit.

\subsection{Heavy meson decay constant}
\label{sec:HQSf}

As a final example of the utility of the heavy quark limit, consider the coupling of the heavy meson field to the axial vector current.  This is conventionally parameterized in terms of a decay constant; for example, for the $B^-$ meson we define $f_B$ via
\be\label{fBdef}
  \langle 0|\,\bar u\gamma^\mu\gamma^5 b\,|B^-(p_B)\rangle=if_Bp_B^\mu\,.
\ee
What is the dependence of the nonperturbative quantity $f_B$ on $m_B$?  To address this question, we rewrite Eq.~(\ref{fBdef}) in a form appropriate to taking the heavy quark limit, $m_B\to\infty$ (which is equivalent to $m_b\to\infty$).  This entails making explicit the dependence of all quantities on $m_B$.  First, we trade the $B^-$ momentum for its velocity,
\be
  p_B^\mu=m_B v^\mu\,.
\ee
Second, we replace the usual $B^-$ state, whose normalization depends on $m_B$,
\be
  \langle B(p_1)|B(p_2)\rangle=2E_B\,\delta^{(3)}(\vec{p}_1-\vec{p}_2)\,,
\ee
by a mass-independent state,
\be
  |B(v)\rangle={1\over\sqrt{m_B}}\,|B(p_B)\rangle\,,
\ee
satisfying
\be
  \langle B(v_1)|B(v_2)\rangle=2\gamma\,\delta^{(3)}(\vec{p}_1-\vec{p}_2)\,.
\ee
Then Eq.~(\ref{fBdef}) becomes
\be
  \sqrt{m_B}\langle 0|\,\bar u\gamma^\mu\gamma^5 b\,|B^-(v)\rangle
  =if_Bm_Bv^\mu\,.
\ee
The nonperturbative matrix element $\langle 0|\,\bar u\gamma^\mu\gamma^5 b\,|B^-(v)\rangle$ is independent of $m_B$ in the heavy quark limit.  Hence, we see that in this limit $f_B$ takes the form
\be
  f_B=m_B^{-1/2}\times{\rm (independent\ of\ }m_B{\rm )}\,.
\ee
This makes explicit the scaling of $f_B$ with $m_B$.  It is more interesting to write this as a prediction for the ratio of charmed and bottom meson decay constants.  We find~\cite{IW,VS,PW}
\be\label{fBfD}
  {f_B\over f_D}=\sqrt{m_D\over m_B} 
  + O\left({\lqcd\over m_D},{\lqcd\over m_B}\right)\,.
\ee
For the physical bottom and charm masses, of course, the correction terms proportional to $\lqcd/m_Q$ could be important.

\section{Heavy Quark Effective Theory}

We have already extracted quite a bit of nontrivial information from the heavy quark
limit.  We have found the scaling of various quantities with $m_Q$, we have studied
the implications for heavy hadron spectroscopy, and we have found nonperturbative
relations among the hadronic form factors which describe semileptonic $b\to c$
decay.  However, all of these results have been obtained in the strict limit
$m_Q\to\infty$.  If the heavy quark limit is to be of more than academic interest,
and is to provide the basis for quantitative phenomenology, we have to understand how
to include corrections systematically.  There are actually two types of corrections
which we would like to include.  {\it Power corrections\/} are subleading terms in
the expansion in $\lqcd/m_Q$; those proportional to $\lqcd/m_c$ are the most
worrisome, because of the relatively small charm quark mass.  {\it Logarithmic
corrections\/} arise from the implicit dependence of quantities on $m_Q$ through the
strong coupling constant $\alpha_s(m_Q)\sim1/\ln(m_Q/\lqcd)$.  For the physical
values of $m_b$ and $m_c$, either of these could be important.  What we need is a
formalism which can accommodate them both.

In short, we need to go from a set of heavy quark symmetry predictions in the
$m_Q\to\infty$ limit, to a reformulation of QCD which provides a controlled
expansion about this limit.  The formalism which does the job is the Heavy Quark
Effective Theory, or the HQET.  The purpose of the HQET is to allow us to extract,
explicitly and systematically, all dependence of physical quantities on $m_Q$, in
the limit $m_Q\gg\lqcd$.  In these lectures, we will develop only enough of the
technology to treat the dominant leading effects, providing indications along the way of
how one would carry the expansion further.

The HQET, as formulated here, was developed in a series of papers going back to the late 1980's~\cite{IW,VS,PW,EH,Georgi,Grin,FGGW,FG,Luke,FGL,MRR}, which the reader who is interested in tracking its historical development may consult.

\subsection{The effective Lagrangian}

Consider the kinematics of a heavy quark $Q$, bound in a hadron with light degrees
of freedom to make a color singlet state.  The small momenta which $Q$ typically exchanges with the rest of the
hadron are of order $\lqcd\ll m_Q$, and they never take $Q$ far from its mass shell,
$p_Q^2=m_Q^2$.  Hence the momentum $p_Q^\mu$ can be decomposed into two parts,
\be
  p_Q^\mu=m_Q v^\mu+k^\mu\,,
\ee
where $m_Qv^\mu$ is the constant on-shell part of $p_Q^\mu$, and $k^\mu\sim\lqcd$ is
the small, fluctuating ``residual momentum''.  The on-shell condition for the heavy
quark then becomes
\be
  m_Q^2=(m_Qv^\mu+k^\mu)^2=m_Q^2+2m_Qv\cdot k+k^2\,.
\ee
In the heavy quark limit we may neglect the last term, and we have the simple
condition
\be
  v\cdot k=0
\ee
for an on-shell heavy quark.  Here the velocity $v^\mu$ functions as a label; since soft interactions cannot change $v^\mu$, there is a {\it velocity superselection rule\/} in the heavy quark limit, and $v^\mu$ is a good quantum number of the QCD Hamiltonian.

We find the same result by taking the $m_Q\to\infty$ limit of the heavy quark propagator,
\be
  {i\over\rlap/p-m_Q+i\epsilon}\to{1+\rlap/v\over2}\,
  {i\over v\cdot k+i\epsilon}\,.
\ee
In this limit the propagator is independent of $m_Q$.  The projection operators
\be
  P_\pm={1\pm\rlap/v\over2}
\ee
project onto the positive ($P_+$) and negative ($P_-$) frequency parts of the Dirac field $Q$.  This is clear in the Dirac representation in the rest frame, in which $P_+$ and $P_-$ project, respectively, onto the upper two and lower two components of the heavy quark spinor.  In the limit $m_Q\to\infty$, in which $Q$ remains almost on shell, only the ``large'' upper components of the field $Q$ propagate; mixing via zitterbewegung with the ``small'' lower components is suppressed by $1/2m_Q$.  Hence the action of the projectors on $Q$ is
\be
  P_+Q(x)=Q(x)+O(1/m_Q)\,,\qquad P_-Q(x)=0+O(1/m_Q)\,.
\ee

The momentum dependence of the field $Q$ is given by its action on a heavy quark state,
\be 
  Q(x)\,|Q(p)\rangle=e^{-ip\cdot x}\,|0\rangle\,.
\ee
If we now multiply both sides by a phase corresponding to the on-shell momentum,
\be 
  e^{im_Qv\cdot x}Q(x)\,|Q(p)\rangle=e^{-ik\cdot x}\,|0\rangle\,,
\ee
the right side of this equation is independent of $m_Q$.  Hence the left side must be, as well.  Combining this observation with the argument of the previous paragraph, we are motivated to define a $m_Q$-{\it independent\/} effective heavy quark field $h_v(x)$,
\be
  h_v(x)=e^{im_Qv\cdot x}\,P_+\,Q(x)\,.
\ee
Note that the effective field carries a velocity label $v$ and is a two-component object.  The modifications to the ordinary field $Q(x)$ project out the positive frequency part and ensure that states annihilated by $h_v(x)$ have no dependence on $m_Q$.  Hence, these are reasonable candidate fields to carry representations of the heavy quark symmetry.
Of course, the small components cannot be neglected when effects of order $1/m_Q$ are included.  In the HQET they are represented by a field
\be
  H_v(x)=e^{im_Qv\cdot x}\,P_-\,Q(x)\,.
\ee
The field $H_v(x)$ vanishes in the $m_Q\to\infty$ limit.

The ordinary QCD Lagrange density for a field $Q(x)$ is given by
\be
  {\cal L}_{\rm QCD}=\overline Q(x)\,(i\Dslash-m_Q)\,Q(x)\,,
\ee
where $D_\mu=\partial_\mu-igA^a_\mu T^a$ is the gauge covariant derivative.  To find the Lagrangian of the HQET, we substitute
\be
  Q(x)=e^{-im_Qv\cdot x}h_v(x)+\dots
\ee
into ${\cal L}_{\rm QCD}$ and expand.  With the aid of the projection identity $P_+\gamma^\mu P_+=v^\mu$, we find
\be
  \lhqet= \bar h_v(x)iv\cdot Dh(x)\,.
\ee
This simple Lagrangian leads to the propagator we derived earlier,
\be
  {i\over v\cdot k+i\epsilon}\,,
\ee
and to an equally simple quark-gluon vertex,
\be
  igT^av^\mu A^a_\mu\,.
\ee
Note that both the propagator and the vertex are independent of $m_Q$, reflecting the heavy quark flavor symmetry.  They also have no Dirac structure, reflecting the heavy quark spin symmetry.  Our intuitive statements about the structure of heavy hadrons have been promoted to explicit symmetries of the QCD Lagrangian in the limit $m_Q\to\infty$.

It is straightforward to include power corrections to $\lhqet$.  Write $Q(x)$ in terms of the effective fields,
\be
  Q(x)=e^{-im_Qv\cdot x}\left[h_v(x)+H_v(x)\right]\,,
\ee
and apply the classical equation of motion $(i\Dslash-m_Q)Q(x)=0$:
\be
  i\Dslash\,h_v(x)+(i\Dslash-2m_Q)H_v(x)=0\,.
\ee
Multiplying by $P_-$ and commuting $\rlap/v$ to the right, we find
\be
  (iv\cdot D+2m_Q)H_v(x)=i\Dslash_\perp\,h_v(x)\,,
\ee
where $D^\mu_\perp=D^\mu-v^\mu v\cdot D$.  We then substitute $Q(x)$ into ${\cal L}_{\rm QCD}$ as before, eliminate $H_v(x)$ and expand in $1/m_Q$ to obtain
\bea\label{hqetlagrangian}
  \lhqet&=&\bar h_viv\cdot Dh_v+\bar h_vi\Dslash_\perp
  \,{1\over iv\cdot D+2m_Q}\,i\Dslash_\perp h_v\nonumber\\
  &=&\bar h_viv\cdot Dh_v+{1\over2m_Q}\left[\bar h_v(iD_\perp)^2h_v
  +{g\over2}\,\bar h_v\sigma^{\alpha\beta}G_{\alpha\beta}h_v\right]
  +\dots\,.
\eea
The leading corrections have a simple interpretation, which becomes clear in the rest frame, $v^\mu=(1,0,0,0)$.  The spin-independent term is
\be
  {1\over2m_Q}\,{\cal O}_K={1\over2m_Q}\,\bar h_v(iD_\perp)^2h_v\to 
  -{1\over2m_Q}\,\bar h_v(i\vec D)^2h_v\,,
\ee
which is the negative of the nonrelativistic kinetic energy of the heavy quark.  Because of the explicit factor of $1/2m_Q$, this term violates the heavy flavor symmetry.  The spin-dependent part is
\be
  {1\over2m_Q}\,{\cal O}_G=
  {1\over2m_Q}\,{g\over2}\,\bar h_v\sigma^{\alpha\beta}G_{\alpha\beta}h_v\to
  {1\over4m_Q}\,\bar h_v\sigma^{ij}T^ah_v\times gG_{ij}^a=
  g\vec\mu_Q^{\,a}\cdot\vec B^{\,a}\,,
\ee
which is the coupling of the spin of the heavy quark to the chromomagnetic field.  Because it has a nontrivial Dirac structure, this term violates both the heavy flavor symmetry and the heavy spin symmetry.  For example, ${\cal O}_G$ is responsible for the $D-D^*$ and $B-B^*$ mass splittings.  These correction terms will be treated as part of the {\it interaction\/} Lagrangian, even though ${\cal O}_K$ has a piece which is a pure bilinear in the heavy quark field.

\subsection{Effective currents and states}

The expansion of the weak interaction current $\bar c\gamma^\mu(1-\gamma^5)b$ is analogous.  However, here we must introduce separate effective fields for the charm and bottom quarks, each with its own velocity:
\be
  b\to h^b_v\,,\qquad c\to h^c_{v'}\,.
\ee
Then a general flavor-changing current becomes, to leading order,
\be
  \bar c\,\Gamma\, b\to\bar h^c_{v'}\,\Gamma\, h^b_v\,,
\ee
where $\Gamma$ is a fixed Dirac structure.  With the leading power corrections, this is
\be\label{current}
  \bar c\,\Gamma\, b\to\bar h^c_{v'}\,\Gamma\, h^b_v
  +{1\over2m_b}\,\bar h^c_{v'}\Gamma(i\Dslash_\perp)h^b_v
  +{1\over2m_c}\,\bar h^c_{v'}(-i\overleftarrow\Dslash_\perp)\Gamma h^b_v
  +\dots\,.
\ee
The effective currents, and other operators which appear in the HQET, may often be simplified by use of the classical equation of motion,
\be
  iv\cdot D h_v(x)=0\,.
\ee
However, it is only safe to apply these equations naively at order $1/m_Q$; at higher order the application of the equations of motion involves additional subtleties~\cite{Pol,FN,FLS94}.

To complete the effective theory, we need $m_Q$-independent hadron states which are created and annihilated by currents containing the effective fields.  For example, there is an effective pseudoscalar meson state $|M(v)\rangle$ which couples to the effective axial current $\bar q\gamma^\mu\gamma^5 h_v$, with a coupling $F_M$ which is independent of $m_Q$,
\be\label{matfm}
  \langle0|\,\bar q\gamma^\mu\gamma^5 h_v\,|M(v)\rangle=iF_Mv^\mu\,.
\ee
At lowest order, $F_M$ is related to a conventional decay constant such as $f_B$ by 
\be
  f_B=F_M/\sqrt{m_B}\,,
\ee
from which we immediately find the relationship (\ref{fBfD}) between $f_D$ and $f_B$.

\begin{figure}
\epsfxsize=12cm
\hfil\epsfbox{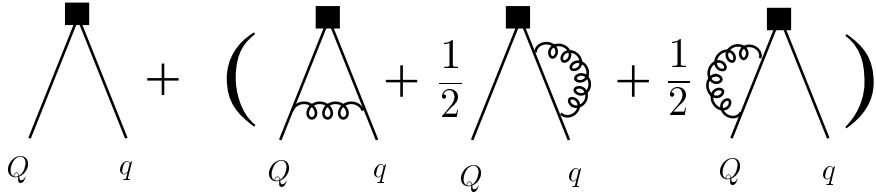}\hfill
\caption{Tree level plus the one loop renormalization of the current $\bar q\Gamma Q$ in QCD.  The box represents the current insertion.}
\label{fig:fqcd}
\end{figure}

\subsection{Radiative corrections}

We can use the effective Lagrangian $\lhqet$ to compute the radiative corrections to the matrix element (\ref{matfm}).  In particular, we would like to extract the dependence of $F_M$ on $\ln m_Q$.  This dependence comes through the one-loop renormalization of the current $\bar q\gamma^\mu\gamma^5Q$.  At lowest order, of course, the renormalization is straightforward: we simply compute the set of graphs found in Fig.~\ref{fig:fqcd}.
The result is finite, because the current is (partially) conserved, and we extract a result of the form
\be\label{qcdcurrent}
  \bar q\gamma^\mu\gamma^5 Q\,\left(1-\bar\gamma_0{\alpha_s\over4\pi}\,
  \ln(m_Q/m_q)+\dots\right)\,.
\ee
Note that there is no explicit dependence on the renormalization scale $\mu$, since there is no divergence to be subtracted.

The same result may be obtained in the effective theory.  In this case we must match the currents in full QCD onto HQET currents of the form $\bar q\gamma^\mu\gamma^5 h_v$.  This step will induce a matching coefficient containing the explicit dependence on $m_Q$, which is absent, by construction, from the operators and Lagrangian of the HQET.\footnote{In general, the matching procedure at order $\alpha_s$ can also induce new Dirac structures $\bar q\,\Gamma\, h_v$.  They do not affect the leading logarithms discussed here.}  In addition, the effective current will not necessarily be conserved, since the ultraviolet properties of QCD and the HQET differ.  Hence the form of the matching, once radiative corrections are included, is
\be
  \bar q\gamma^\mu\gamma^5 Q\to C(m_Q,\mu)\,\bar q\gamma^\mu\gamma^5 h_v\,.
\ee

\begin{figure}
\epsfxsize=13cm
\hfil\epsfbox{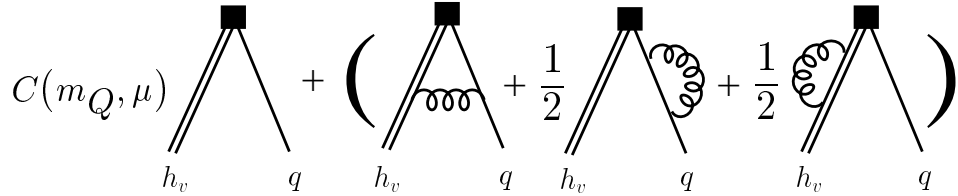}\hfill
\caption{Tree level plus the one loop renormalization of the current $\bar q\gamma^\mu\gamma^5 h_v$ in HQET.  The double line is the propagator of the effective field $h_v$.}
\label{fig:fhqet}
\end{figure}

We can deduce the form of $C(m_Q,\mu)$ by considering the renormalization of the effective current $\bar q\gamma^\mu\gamma^5 h_v$, shown in the last three terms of Fig.~\ref{fig:fhqet}.  These diagrams, computed in the effective theory, are independent of $m_Q$.  However, in general they are divergent, so they depend on the renormalization scale $\mu$; the renormalization takes the form
\be
  \bar q\gamma^\mu\gamma^5 h_v\,\left(C(m_Q,\mu)-\gamma_0{\alpha_s\over\pi}\,
  \ln(\mu/m_q)+\dots\right)\,,
\ee
where here $m_q$ acts as an infrared cutoff.  The $\mu$ dependence in the second term must be canceled by $C(m_Q,\mu)$.  Since the logarithm depends on a dimensionless ratio, $C(m_Q,\mu)$ must be of the form
\be
  C(m_Q,\mu)=1-\gamma_0{\alpha_s\over\pi}\,\ln(m_Q/\mu)+\dots\,.
\ee
Comparing the dependence on $m_Q$ of $C(m_Q,\mu)$ and the expansion (\ref{qcdcurrent}), we see that $\bar\gamma_0=\gamma_0$.

However, the effective theory allows us to go beyond leading order, and to resum all corrections of the form $\alpha_s^n\ln^n m_Q$.  We do this with the renormalization group equations, which express the independence of physical observables on the renormalization scale $\mu$.  In this case, they require that the $\mu$ dependence of $C(m_Q,\mu)$ cancel that of the one loop diagrams in Fig.~\ref{fig:fhqet}, under small changes in $\mu$:
\be
  \mu{{\rm d}\over{\rm d}\mu}\,C(m_Q,\mu)
  =\gamma_0\,{\alpha_s(\mu)\over\pi}\,.
\ee
Here the total derivative with respect to $\mu$ includes the implicit dependence on $\mu$ of the coupling constant $\alpha_s(\mu)$:
\bea
  &&\mu{{\rm d}\over{\rm d}\mu}=\mu{\partial\over\partial\mu}
  +\beta(g){\partial\over\partial g}\nonumber\\
  &&\beta(g)=-\beta_0{g^3\over16\pi^2}+\dots\nonumber\\
  &&\beta_0=11-{2\over3}N_f={25\over3}{\rm \ \ for\ }N_f=4\,,
\eea
where $N_f$ is the number of light flavors.  We compute the anomalous dimension $\gamma_0$ from the divergent parts of the one loop diagrams shown in Fig.~\ref{fig:fhqet}.  From the three terms we find, respectively~\cite{VS,PW},
\be
  \gamma_0={8\over3}+{1\over2}\left(-{8\over3}\right)
  +{1\over2}\left({16\over3}\right)=4\,.
\ee
The solution of the renormalization group equation is
\be
  C(m_Q,\mu)=\left({\alpha_s(m_Q)\over\alpha_s(\mu)}\right)
  ^{\gamma_0/2\beta_0}
  =\left({\alpha_s(m_Q)\over\alpha_s(\mu)}\right)^{6/25}\,.
\ee
This then yields the leading logarithmic correction to the ratio $f_B/f_D$:
\be
  {f_B\over f_D}=\sqrt{{m_D\over m_B}}
  \left({\alpha_s(m_b)\over\alpha_s(m_c)}\right)^{6/25}\,.
\ee
The radiative correction is approximately a ten percent effect.  In fact, it has a simple physical interpretation.  For virtual gluons of ``intermediate'' energy, $m_c<E_g<m_b$, the bottom quark is heavy but the charm quark is light.  Such gluons contribute to the difference between $f_B$ and $f_D$ even in the heavy quark limit.

In summary, then, the purpose of the HQET is to make explicit all dependence of observable quantities on $m_Q$.  The logarithmic dependence, through $\alpha_s(m_Q)$, arises from intermediate virtual gluons with $m_c<E_g<m_b$.  We obtain these corrections by computing perturbatively with the HQET Lagrangian, then using the renormalization group to resum the logarithms to all orders.  The power dependence, $1/m_Q$, is extracted systematically in the heavy quark expansion.  We have seen how to expand the Lagrangian and the states to subleading order; the application of the expansion to a physical decay rate will be presented in the next section.

These lectures are meant to be pedagogical, so we will only treat the leading corrections to a few processes.  However, the state of the art goes significantly beyond what will be presented here.  For many quantities, not only the leading logarithms, $\alpha_s^n\ln^n m_Q$, but the subleading (two loop) logarithms, of order $\alpha_s^{n+1}\ln^n m_Q$, have been resummed.  Similarly, many power corrections are known to relative order $1/m_Q^2$.  It is particularly important phenomenologically to take into account the corrections of order $1/m_c^2$.

\section{Exclusive $B$ Decays}

We now have the tools we need for an HQET treatment of the exclusive semileptonic transitions $B\to D\,\ell\bar\nu$ and $B\to D^*\,\ell\bar\nu$.  Earlier, we argued on physical grounds that in the heavy quark limit all of the hadronic matrix elements which appear in these decays are related to a single nonperturbative function $\xi(w)$.  Now we will sharpen this analysis to actually {\it derive\/} these relations, and to include radiative and power corrections.  In fact, almost all of our effort will go into the power corrections, since the radiative corrections to the transition currents are computed just as in the previous section.

\subsection{Matrix element relations at leading order}

The transitions in question require the nonperturbative matrix elements
\be
  \langle D(v')|\,\bar c\gamma^\mu b\,|B(v)\rangle\,,\qquad
  \langle D^*(v',\epsilon)|\,\bar c\gamma^\mu b\,|B(v)\rangle\,,\qquad
  \langle D^*(v',\epsilon)|\,\bar c\gamma^\mu\gamma^5 b\,|B(v)\rangle\,,
\ee
parameterized in terms of form factors as in Eq.~(\ref{formfactors}).  Our first task is to derive the relations between these form factors, as promised earlier.  These relations depend on the heavy quark symmetry, that is, on the fact that the spin quantum numbers of $Q$ and of the light degrees of freedom are separately conserved by the soft physics.  Hence we need a representation of the heavy meson states in which they have well defined transformations separately under the angular momentum operators $S_Q$ and $J_\ell$.  In particular, the representation must reflect the fact that a rotation by $S_Q$ can exchange the pseudoscalar meson $M(v)$ with the vector meson $M^*(v,\epsilon)$.

The solution is to introduce a ``superfield'' ${\cal M}(v)$, defined as the $4\times4$ Dirac matrix~\cite{FGGW,Falk92}
\be
  {\cal M}(v)={1+\rlap/v\over2}\left[\gamma^\mu M^*_\mu(v,\epsilon)
  -\gamma^5 M(v)\right]\,.
\ee
Under heavy quark spin rotations $S_Q$, ${\cal M}(v)$ transforms as
\be\label{Mtransform}
  {\cal M}(v)\to D(S_Q){\cal M}(v)\,,
\ee
and under Lorentz rotations $\Lambda$, as
\be
  {\cal M}(v)\to D(\Lambda){\cal M}(\Lambda^{-1}v)D^{-1}(\Lambda)\,.
\ee
Here $D(\cdots)$ is the spinor representation of $SO(3,1)$.  The superfield satisfies the matrix identity
\be\label{Midentity}
  P_+\,{\cal M}(v)\,P_-={\cal M}(v)\,,
\ee
so it transforms the same way as the product of spinors $h_v\,\bar q$, representing a heavy quark and a light antiquark moving together at velocity $v^\mu$.

A current which mediates the decay of one heavy quark ($Q$) into another ($Q'$) is of the form $\bar h_{v'}\,\Gamma\,h_v$.  Under a rotation by $S_Q$, the effective field $h_v$ transforms as
\be
  h_v\to D(S_Q)\,h_v\,,
\ee
while $h_{v'}$ is unchanged.  The current would remain invariant if we took $\Gamma$ to transform as
\be
  \Gamma\to\Gamma\,D^{-1}(S_Q)\,.
\ee
On the other hand, the matrix element of superfields
\be
  \langle {\cal M}'(v')|\,\bar h_{v'}\,\Gamma\,h_v\,|{\cal M}(v)\rangle
\ee
is invariant if we rotate {\it both\/} $h_v$ and ${\cal M}(v)$ by the {\it same} $S_Q$.  With the transformation law (\ref{Mtransform}) for ${\cal M}(v)$, it follows that the $S_Q$-invariant matrix element must be proportional to $\Gamma{\cal M}(v)$.   When we also consider rotations under $S_{Q'}$, we find that the matrix element is restricted to the general form
\be\label{matelgen}
  \langle {\cal M}'(v')|\,\bar h_{v'}\,\Gamma\,h_v\,|{\cal M}(v)\rangle=
  -\left[M_MM_{M'}\right]^{1/2}\,{\rm Tr}\,
  \left[\overline{\cal M}'(v')\,\Gamma\,{\cal M}(v)\,\hat F(v,v')\right]\,.
\ee
The product of masses in front is a convention which restores the relativistic normalization of the states.  Note that the heavy quark symmetry allows an arbitrary $4\times4$ Dirac matrix $\hat F(v,v')$ to act on the ``light quark'' part of the superfields.  Its presence reflects the fact that, other than the constraint of Lorentz symmetry, the behavior of the spin of the light degrees of freedom during the decay is unknown.

A general expansion of $\hat F(v,v')$ in terms of scalar functions $F_i(w=v\cdot v')$ takes the form
\be
  \hat F(v,v')=F_1(w)+F_2(w)\rlap/v+F_3(w)\rlap/v'+F_4(w)\rlap/v\rlap/v'\,.
\ee
However, the identity (\ref{Midentity}) applied to the matrix element (\ref{matelgen}) yields
\be
  \hat F(v,v')=P_-\,\hat F(v,v')\,P'_-=\left[F_1(w)-F_2(w)-F_3(w)+F_4(w)
  \right]P_-\,P'_-\,.
\ee
In other words, $F(v,v')$ is actually a {\it scalar\/}, which we identify with the Isgur-Wise function,
\be
   \hat F(v,v')=\xi(w)\,.
\ee
As an exercise, let us apply this formalism to the matrix elements for $B\to(D,D^*)\,\ell\bar\nu$.  For a given matrix element, we pick out the part of the superfield ${\cal M}(v)$ which is relevant.  Hence we find
\bea
 \langle D(v')|\,\bar c\gamma^\mu b\,|B(v)\rangle &=&
  \langle M_D(v')|\,\bar h^c_{v'}\gamma^\mu h^b_v\,|M_B(v)\rangle\nonumber\\
  &=& -\left[m_Dm_B\right]^{1/2}\,{\rm Tr}\,
  \left[\gamma^5P'_+\gamma^\mu P_+(-\gamma^5)\right]\,\xi(w)\nonumber\\
  &=&\left[m_Dm_B\right]^{1/2}\,\xi(w)\,(v+v')^\mu\\ \nonumber\\
 \langle D^*(v',\epsilon)|\,\bar c\gamma^\mu\gamma^5 b\,|B(v)\rangle &=&
  \langle M^*_D(v',\epsilon)|\,\bar h^c_{v'}\gamma^\mu\gamma^5 h^b_v\,
  |M_B(v)\rangle\nonumber\\
  &=&   -\left[m_{D^*}m_B\right]^{1/2}\,{\rm Tr}\,
  \left[\rlap/\epsilon P'_+\gamma^\mu\gamma^5
   P_+(-\gamma^5)\right]\,\xi(w)\nonumber\\
  &=&\left[m_{D^*}m_B\right]^{1/2}\,\xi(w)\,\left[(w+1)\epsilon^{*\mu}
  -\epsilon^*\cdot(v+v')^\mu\right]\\ \nonumber\\
 \langle D^*(v',\epsilon)|\,\bar c\gamma^\mu b\,|B(v)\rangle &=&
  \langle M^*_D(v',\epsilon)|\,\bar h^c_{v'}\gamma^\mu h^b_v\,
  |M_B(v)\rangle\nonumber\\
  &=&   -\left[m_{D^*}m_B\right]^{1/2}\,{\rm Tr}\,
  \left[\rlap/\epsilon P'_+\gamma^\mu P_+(-\gamma^5)\right]\,\xi(w)\nonumber\\
  &=&\left[m_{D^*}m_B\right]^{1/2}\,\xi(w)\,
  i\varepsilon^{\mu\nu\alpha\beta}\epsilon_\nu^*v'_\alpha v_\beta\,,
\eea
reproducing explicitly the relations (\ref{ffrelations}) between the independent form factors $h_i(w)$.  We can also derive the normalization condition at $w=1$.  Consider the matrix element of the $b$ number current $\bar b\gamma^\mu b$ between $B$ meson states.  In QCD, the matrix element of this current is exactly normalized,
\be
  \langle B(v)|\,\bar b\gamma^\mu b\,|B(v)\rangle=2p_B^\mu=2m_B v^\mu\,.
\ee
But in HQET, we have
\bea
  \langle B(v)|\,\bar b\gamma^\mu b\,|B(v)\rangle&=&
  \langle M_B(v)|\,\bar h_v\gamma^\mu h_v\,|M_B(v)\rangle\nonumber\\
  &=& m_B\xi(v\cdot v)(v+v)^\mu\nonumber\\
  &=&2m_Bv^\mu\,\xi(1)\,.
\eea
Hence the normalization condition at zero recoil,
\be
  \xi(1)=1\,,
\ee
follows directly from the conservation of the heavy quark number current.

\subsection{Power corrections to the matrix elements}

The matrix elements we have derived are computed in the strict limit $m_{b,c}\to\infty$.  How are they affected by corrections of order $1/m_b$ and $1/m_c$?  There are two sources of $1/m_Q$ corrections in the effective theory: the corrections (\ref{current}) to the heavy quark currents, and the corrections (\ref{hqetlagrangian}) to the Lagrangian.

When radiative corrections are included, the expansion of the heavy quark current $\bar c\Gamma b$  in terms of HQET operators has a form which is somewhat more general than Eq.~(\ref{current}),
\be
  \bar c\Gamma b\to a_0(\alpha_s)\,\bar h^c_{v'}\Gamma h^b_v
  +{a_1(\alpha_s)\over2m_b}\,\bar h^c_{v'}\Gamma_1^\alpha iD_\alpha h^b_v
  +{a_1'(\alpha_s)\over2m_c}\,\bar h^c_{v'}
  (-i\overleftarrow D_\alpha){\Gamma_1'}^\alpha h^b_v
  +\dots\,.
\ee
The matrix elements of the power corrections are constrained by heavy quark symmetry in a manner completely analogous to the leading current.  In terms of traces over the superfields, we have~\cite{Luke}
\be
  \langle {\cal M}'(v')|\,\bar h_{v'}\,\Gamma^\alpha iD_\alpha\,h_v\,
  |{\cal M}(v)\rangle=
  -\left[M_MM_{M'}\right]^{1/2}\,{\rm Tr}\,
  \left[\overline{\cal M}'(v')\,\Gamma^\alpha\,{\cal M}(v)\,
  \hat G_\alpha(v,v')\right]\,,
\ee
where $\hat G_\alpha(v,v')$ is another arbitrary $4\times4$ Dirac matrix.  The matrix element 
\be
  \langle {\cal M}'(v')|\,\bar h_{v'}\,
  (-i\overleftarrow D_\alpha){\Gamma_1'}^\alpha h_v\,|{\cal M}(v)\rangle
\ee
may also be written in terms of $\hat G_\alpha(v,v')$, using charge conjugation.

The $1/m_Q$ corrections ${\cal O}_K$ and ${\cal O}_G$ to the Lagrangian contribute somewhat differently.   In order to apply heavy quark symmetry, the matrix elements of the local currents, both leading and subleading, must be written in terms of the {\it effective\/} states $|M(v)\rangle$.  However, these states are not eigenstates of the Hamiltonian, once ${\cal O}_K$ and ${\cal O}_G$ are included in the Lagrangian.  Hence we must allow for the possibility that if an effective state $|M(v)\rangle$ is created at time $t=-\infty$, then ${\cal O}_K$ or ${\cal O}_G$ could act on the state before its decay at $t=0$.  This possibility is accounted for by including time-ordered products in which ${\cal O}_K$ or ${\cal O}_G$ is inserted along the incoming or outgoing heavy quark line.  If we are keeping terms of order $1/m_Q$, only one insertion of ${\cal O}_K$ or ${\cal O}_G$ needs to be included.  The time-ordered products are of the form~\cite{Luke}
\bea
  \langle D^{(*)}(v)|\,\bar c\Gamma b\,|B(v)\rangle = \dots &+&
  {1\over2m_c}\,\langle {\cal M}'(v')|\,i\int{\rm d}y\,T\left\{
  \bar h^c_{v'}\Gamma h^b_v,{\cal O}^{v'}_K+{\cal O}^{v'}_G\right\}
  |{\cal M}(v)\rangle\nonumber\\
  &+&{1\over2m_b}\,\langle {\cal M}'(v')|\,i\int{\rm d}y\,T\left\{
  \bar h^c_{v'}\Gamma h^b_v,{\cal O}^v_K+{\cal O}^v_G\right\}
  |{\cal M}(v)\rangle\,,
\eea
where the ellipses refer to the current corrections computed earlier.  The evaluation of the matrix elements of the time-ordered products will lead to still more nonperturbative functions like $\hat F(v,v')$ and $\hat G_\alpha(v,v')$.

\subsection{Corrections at zero recoil}

It is straightforward, but not very illuminating, to expand all of the new nonperturbative functions which arise at order $1/m_Q$ in terms of scalar form factors.  In the end, the corrections may be parameterized in terms of four functions of the velocity transfer $w$, and a single nonperturbative parameter $\bar\Lambda$, all proportional to the mass scale $\lqcd$.  The new parameter has a simple interpretation as the ``energy'' of the light degrees of freedom, and is given by
\be\label{lambardef}
  \bar\Lambda=\lim_{m_b\to\infty}\,(m_B-m_b)\,.
\ee

Instead of a general treatment, however, we will consider the $1/m_Q$ corrections at the zero recoil point $w=1$.  This is clearly the most important case, because it is at this point that the nonperturbative matrix elements are absolutely normalized in the heavy quark limit.  What happens to this normalization condition when $1/m_Q$ corrections are included?

Let us study the corrections to the current in detail.  They are described by the nonperturbative function $\hat G_\alpha(v,v')$.  At $v=v'$, $\hat G_\alpha(v,v)$ may be expanded as
\be
  \hat G_\alpha(v,v)=G_1 v_\alpha+G_2 \gamma_\alpha+G_3 v_\alpha\rlap/v
  +G_4 \gamma_\alpha\rlap/v\,.
\ee
But $\hat G_\alpha(v,v)$ is subject to the same constraint as $\hat F(v,v')$,
\be
  \hat G_\alpha(v,v)=P_-\hat G_\alpha(v,v)P_-=(G_1-G_2-G_3+G_4)v_\alpha P_-
  \equiv G\,v_\alpha P_-\,,
\ee
and it, too, is equivalent to a Dirac scalar (the same is {\it not\/} true of the general function $\hat G_\alpha(v,v')$).  Now consider the matrix element where we take $\Gamma^\alpha=v^\alpha$.  Then we have
\bea
  \langle {\cal M}'(v)|\,\bar h_{v'}\,iv\cdot D\,h_v\,
  |{\cal M}(v)\rangle
  &=&\mbox{}-\left[M_MM_{M'}\right]^{1/2}\,{\rm Tr}\,
  \left[\overline{\cal M}'(v)\,v^\alpha\,{\cal M}(v)\right]\,
  G\,v_\alpha\nonumber\\
  &=& \mbox{}-G\left[M_MM_{M'}\right]^{1/2}\,{\rm Tr}\,
  \left[\overline{\cal M}'(v)\,{\cal M}(v)\right]\,.
\eea
But this matrix element {\it vanishes\/} by the classical equation of motion in the effective theory,
\be
  v\cdot D\,h_v(x)=0\,.
\ee
Hence $G=0=\hat G_\alpha(v,v)$.  There are no $1/m_Q$ corrections from the current to the normalization condition at zero recoil~\cite{Luke,CG}.

The same is true of insertions of the corrections ${\cal O}_K$ and ${\cal O}_G$ to the Lagrangian: their contribution vanishes at $w=1$.  To show this requires the imposition of the conservation of the $b$ number current at order $1/m_b$, much as we derived the normalization of the Isgur-Wise function at leading order.  The complete derivation is found in the literature~\cite{Luke}.

In the end, we have an extremely exciting result, known as {\it Luke's Theorem.}  There are no corrections at zero recoil to the hadronic matrix elements responsible for the semileptonic decays $B\to D\,\ell\bar\nu$ and $B\to D^*\,\ell\bar\nu$.  The leading power corrections to the normalization of zero recoil matrix elements are only of order $1/m_c^2$.  Given that $\lqcd/m_c\sim30\%$ and $\lqcd^2/m_c^2\sim10\%$, the implication is that the leading order predictions at $w=1$ are considerably more accurate than one might have expected.  In addition, away from zero recoil the $1/m_c$ corrections must be suppressed at least by $(w-1)$.

On closer inspection, this result is more interesting for $B\to D^*\,\ell\bar\nu$ than for $B\to D\,\ell\bar\nu$.  This is because the leading order matrix element for $B\to D\,\ell\nu$ vanishes kinematically at zero recoil for a massless lepton in the final state.  Hence, in this case the $1/m_c$ corrections are not suppressed as a fractional correction to the lowest order term~\cite{NR}.

\subsection{Extraction of $|V_{cb}|$ from $B\to D^*\,\ell\bar\nu$}

An immediate application of these results is the extraction of $|V_{cb}|$ from the exclusive decay $B\to D^*\,\ell\bar\nu$.  This process is mediated by the weak operator ${\cal O}_{bc}$~(\ref{Obcdef}), whose matrix element factorizes as
\be
  \langle D^*\,\ell\bar\nu|\,{\cal O}_{bc}\,|B\rangle=
  {G_FV_{cb}\over\sqrt2}\,
  \langle D^*|\,\bar c\gamma^\mu(1-\gamma^5)b\,|B\rangle\,
  \langle\ell\bar\nu|\,\bar\ell\gamma_\mu(1-\gamma^5)\nu\,|0\rangle\,.
\ee
The leptonic matrix element may be computed perturbatively, while we treat the 
hadronic matrix element in the heavy quark expansion.  The result is a differential decay rate of the form~\cite{Neu91}
\bea\label{diffrate}
  {{\rm d}\Gamma\over{\rm d}w}&=&{G_F^2\over48\pi^3}\,|V_{cb}|^2\,
  (m_B-m_{D^*})^2m_{D^*}^3(w+1)^3\sqrt{w^2-1}\nonumber\\
  &&\quad\times\left[1+{4w\over w+1}\,{m_B^2-2wm_bm_{D^*}+m_{D^*}^2
  \over(m_B-m_{D^*})^2}\right]F^2(w)\,.
\eea
All of the HQET analysis goes into the factor $F(w)$, which has an expansion
\be
  F(w)=\xi(w)+{\rm (radiative\ corrections)}+
  {\rm (power\ corrections)}\,.
\ee

We extract $|V_{cb}|$ by studying the differential decay rate near $w=1$, where the hadronic matrix elements are known.  Of course, this requires extrapolation of the experimental data, since the rate vanishes kinematically at $w=1$.  For massless leptons, only the matrix element $\langle D^*|\,\bar c\gamma^\mu\gamma^5b\,|B\rangle$ of the axial current contributes at this point.  The analysis of this quantity in the HQET yields an expansion of the form
\be
  F(1) =\eta_A\,\left[1+{0\over m_c}+{0\over m_b}
  +\delta_{1/m^2}+\dots\right]\,.
\ee
The correction $\delta_{1/m^2}$, which contains terms proportional to $1/m_c^2$, $1/m_b^2$ and $1/m_cm_b$, is intrinsically nonperturbative.  It has been estimated from a variety of models to be small and negative~\cite{FN,SUV,Neu94},
\be
  \delta_{1/m^2}\approx -0.055\pm0.035\,.
\ee
Note that the model dependence in the result has been relegated to the estimation of the sub-subleading terms.  The radiative correction $\eta_A$ has now been computed to two loops~\cite{Cz,FGN},
\be
  \eta_A=0.960\pm0.007\,.
\ee
The result is a value for $F(1)$ with errors at the level of 5\%,
\be
  F(1)=0.91\pm0.04\,.
\ee
This is the theory error which the experimental determination of $|V_{cb}|$ will inherit.  It is dominated by the uncertainty in the nonperturbative corrections, and it is difficult to see how this can be improved much in the future.

All that is left experimentally is to extrapolate the data to $w=1$ and extract
\be
  \lim_{w\to1}\,{1\over\sqrt{w-1}}\,{{\rm d}\Gamma\over{\rm d}w}\,.
\ee
Once the kinematic factors in Eq.~(\ref{diffrate}) have been included, this amounts to a direct measurement of the combination $|V_{cb}|F(1)$.  Four experiments recently have reported results for this quantity~\cite{AL1,AR1,CL1,DE1}:
\bea
  {\rm ALEPH:}&\qquad&(31.4\pm2.3\pm2.5)\times10^{-3}\nonumber\\
  {\rm ARGUS:}&\qquad&(38.8\pm4.3\pm2.5)\times10^{-3}\nonumber\\
  {\rm CLEO:}&\qquad&(35.1\pm1.9\pm2.0)\times10^{-3}\nonumber\\
  {\rm DELPHI:}&\qquad&(35.0\pm1.9\pm2.3)\times10^{-3}\,.
\eea
Neubert~\cite{Neu96} has combined these measurements to find a ``world average'' of $(35.3\pm1.8)\times10^{-3}$, from which we extract
\be\label{vcbexcl}
  |V_{cb}|=(38.8\pm2.0_{\rm exp}\pm1.6_{\rm th})\times10^{-3}\,.
\ee
This value of $|V_{cb}|$ has almost no dependence on hadronic models.  In contrast to model-based ``measurements'', here the theoretical error is meaningful, in that it is based on a systematic expansion in small quantities.

\section{Inclusive $B$ Decays}

An exclusive semileptonic $B$ decay, such as $B\to D\,\ell\bar\nu$, is one in which the final hadronic state is fully reconstructed.  An inclusive decay, by contrast, is one in which only certain kinematic features, and perhaps the flavor, of the hadron are known.  In this case, we need a theoretical analysis in which we sum over all possible hadronic final states allowed by the kinematics.  Fortunately, this is possible within the structure of the HQET.

As in the case of exclusive decays, the key theme is the separation of short distance physics, associated with the heavy quark, from long distance physics, associated with the light degrees of freedom.  We will also rely on heavy quark spin and flavor symmetry.  However, the new ingredient will be the idea of ``parton-hadron duality'', which, as we will see, also relies on the heavy quark limit $m_b\gg\lqcd$.

\subsection{The inclusive decay $B\to X_c\,\ell\bar\nu$}

Let us consider the inclusive decay
\be
  B(p_B)\to X_c(p_X)\,\ell(p_\ell)\nu(p_\nu)\,,
\ee
where all that is known about the state $X_c$ is its energy and momentum, and the fact that it contains a charm quark.  This decay is mediated by the weak operator ${\cal O}_{bc}$.  It is easy to generalize our discussion to inclusive decays of other heavy quarks, such as $b\to u\,\ell\nu$ and $c\to s\,\bar\ell\nu$, by replacing ${\cal O}_{bc}$ with the appropriate weak operator.

The treatment of exclusive decays required both the $b$ and $c$ quarks to be heavy.  For inclusive decays we can relax this condition on the $c$ quark, requiring only $m_b\gg\lqcd$.  What does the weak decay of the $b$, at time $t=0$, look like to the light degrees of freedom?  For $t<0$, there is a heavy hadron composed of a point-like color source and light quarks and gluons.  At $t=0$, the point source disappears, releasing both its color and a large amount of energy into the hadronic environment.  Eventually, for $t>0$, this new collection of strongly interacting particles will materialize as a set of physical hadrons.  The probability of this hadronization is unity; there is no interference between the hadronization process and the heavy quark decay.  There are subleading effects in powers of $\lqcd/m_b$, but they do not alter the probability of hadronization.  Rather, they reflect the fact that the $b$ quark is not exactly a static source of color:  it has a small nonrelativistic kinetic energy and it carries a spin, both of which affect the kinematic properties of its decay.

As in the case of exclusive decays, we will compute the inclusive semileptonic width $\Gamma(B\to X_c\,\ell\bar\nu)$ as a double expansion in $\alpha_s(m_b)$ and $\lqcd/m_b$~\cite{FLS94,CGG,SV,MW}.  The expansion in $\alpha_s(m_b)$ reflects the applicability of perturbative QCD to the short distance part of the process.  The heavy quark expansion will be continued to relative order $1/m_b^2$, as there is an analogue of Luke's Theorem which eliminates power corrections to the rate of order $1/m_b$.  The $1/m_b^2$ corrections will be written in terms of three nonperturbative parameters.  The first, $\bar\Lambda$, is defined in Eq.~(\ref{lambardef}).  It is essentially the mass of the light degrees of freedom  in the heavy hadron, but we will see that it is plagued by an ambiguity of order $\lqcd$ in the definition of the $b$ quark mass.  The other two parameters are the expectation values in the $B$ meson of the leading corrections ${\cal O}_K$ and ${\cal O}_G$ to $\lhqet$.  They are defined as~\cite{FN}
\bea
  \lambda_1&=&{1\over2m_B}\langle B|\,{\cal O}_K\,|B\rangle\nonumber\\
  \lambda_2&=&-{1\over6m_B}\langle B|\,{\cal O}_G\,|B\rangle\,,
\eea
where we take the usual relativistic normalization of the states.  Hence, $\lambda_1$ may be thought of roughly as the negative of the $b$ quark kinetic energy, and $\lambda_2$ as the energy of its hyperfine interaction with the light degrees of freedom.

Now let us outline the computation.  The inclusive decay involves a sum over all possible final states, which is actually a sum over exclusive modes (such as $D, D^*,D\pi,\ldots$), followed by a phase space integral for each mode.  We write
\be
  \Gamma(B\to X_c\,\ell\bar\nu)=\sum_{X_c}\int{\rm d}[{\rm P.S.}]\;
  \Big|\langle X_c\,\ell\bar\nu|\,{\cal O}_{bc}\,|B\rangle\Big|^2\,.
\ee
There is an Optical Theorem for QCD, which follows from the analyticity of the scattering matrix as a function of the momenta of the asymptotic states.  Its content is that a transition rate is proportional to the imaginary part of the forward scattering amplitude with two insertions of the transition operator,
\be
  \Gamma(B\to X_c\,\ell\bar\nu)=-2\,{\rm Im}\;i\int{\rm d}x\,e^{ik\cdot x}\,
  \langle B|\,T\left\{{\cal O}^\dagger_{bc}(x),{\cal O}_{bc}(0)\right\}
  \,|B\rangle\equiv 2\,{\rm Im}\,T\,.
\ee
In what follows, we will write the time-ordered product $T\{{\cal O}^\dagger_{bc},{\cal O}_{bc}\}$ as a series of local operators, using the Operator Product Expansion.  As we will see, the applicability of this expansion, and its computation in perturbation theory, will rest on the limit $m_b\gg\lqcd$.  We will then use this limit again to expand the matrix elements of these local operators in the HQET.

The first step is to factorize the integration over the lepton momenta, which can be performed explicitly.  Written as a product of currents, ${\cal O}_{bc}$ takes the form
\be
  {\cal O}_{bc}={G_FV_{cb}\over\sqrt2}\,J^\mu_{bc}\,J_{\ell\mu}\,,
\ee
where
\bea
  J^\mu_{bc}&=&\bar c\gamma^\mu(1-\gamma^5)b\nonumber\\
  J^\mu_\ell&=&\bar\ell\gamma^\mu(1-\gamma^5)\nu\,.
\eea
Then $T$ can be decomposed as an integral over the total momentum $q^\mu=p_\ell^\mu+p_\nu^\mu$ transferred to the leptons,
\be\label{Tdef}
  T={1\over2}G_F^2|V_{cb}|^2\int{\rm d}q\,T^{\mu\nu}(q)\,L_{\mu\nu}(q)\,.
\ee
Here the lepton tensor is
\bea
  L_{\mu\nu}(q)&=&\int{\rm d}[{\rm P.S.}]\;
  \langle0|\,J^\dagger_{\ell\mu}\,|\ell\bar\nu\rangle\,
  \langle\ell\nu|\,J_{\ell\nu}\,|0\rangle\nonumber\\
  &=&{1\over3\pi}\,\left(q_\mu q_\nu-q^2g_{\mu\nu}\right)\,,
\eea
and the hadron tensor is
\be
  T^{\mu\nu}(q)=-i\int{\rm d}x\,e^{iq\cdot x}\,\langle B|\,T\left\{
  J^{\mu\dagger}_{bc}(x),J^\mu_{bc}(0)\right\}\,|B\rangle\,.
\ee
We will need the imaginary part, ${\rm Im}\;T^{\mu\nu}$.  Where is it nonvanishing?  In quantum field theory, a scattering amplitude develops an imaginary part when there can be a real intermediate state, that is, the intermediate particles can all go on their mass shell.  Whether this is possible, of course, depends on the kinematics of the external states.  

\begin{figure}
\epsfxsize=10cm
\hfil\epsfbox{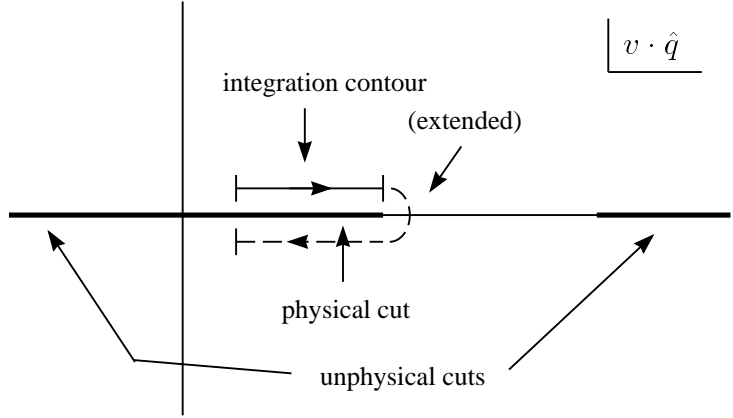}\hfill
\caption{Analytic structure of $T^{\mu\nu}$ in the complex $v\cdot\hat q$ plane, for fixed, real $\hat q^2$.  The integration contour is over the ``physical cut'', corresponding to real decay into leptons.  The unphysical cuts correspond to other processes.}
\label{fig:vqplane}
\end{figure}

In this case, there are two avenues for creating a physical intermediate state~\cite{CGG}.  The first is to act on the external state $|B\rangle$ with the transition current $J^\mu_{bc}$.  The state which is created has no net $b$ number and a single charm quark; the simplest possibility is the decay process $b\to c$.  The momentum of the intermediate state is $p_X=p_B-q$; the condition that it could be on mass shell is simply
\be
  p_X^2=(p_B-q)^2\ge m_D^2\,.
\ee
If we define scaled variables
\be
  p_B^\mu=m_B v^\mu\,,\qquad\hat q^\mu=q^\mu/m_B\,,\qquad
  \hat m_D=m_D/m_B\,,
\ee
this condition becomes
\be
  v\cdot\hat q\le{1\over2}\left(1+\hat q^2-\hat m_D^2\right)\,.
\ee
Another possibility is to act on $|B\rangle$ with the conjugate operator $J^{\mu\dagger}_{bc}$.  This operation would produce an intermediate state with two $b$ quarks and one $\bar c$.  For this state to be on shell, the momentum transfer has to satisfy
\be
  p_X^2=(p_B+q)^2\ge(2m_B+m_D)^2\,,
\ee
that is,
\be
  v\cdot\hat q\ge{1\over2}\left(3-\hat q^2+4\hat m_D+\hat m_D^2\right)\,.
\ee
The physical intermediate states are shown as cuts in the $v\cdot\hat q$ plane in Fig.~\ref{fig:vqplane}.  Also shown is the contour corresponding to the  phase space integration over the lepton momentum $q$.  For physical (massless) leptons which are the product of a heavy quark decay, this integral runs over the top of the lower cut, for the range
\be
  \sqrt{\hat q^2}+i\epsilon\le v\cdot\hat q\le
  {1\over2}\left(1+\hat q^2-\hat m_D^2\right)+i\epsilon\,.
\ee
As indicated by the dotted line, we can continue this contour around the end of the cut and back along the bottom, to $v\cdot\hat q=\sqrt{\hat q^2}-i\epsilon$.  Since $T^{\mu\nu}(v\cdot\hat q\,^*)=-T^{\mu\nu}(v\cdot\hat q)$ for real $\hat q^2$, we compensate for extending the contour by dividing the new integral by two.

We now encounter our central problem.  The integral over $v\cdot\hat q$ runs over physical intermediate hadron states, which are color neutral bound states of quarks and gluons.  Hence the integrand depends intimately on the details of QCD at long distances, which is intrinsically nonperturbative.  A perturbative calculation of $T^{\mu\nu}$, which is all we have at our disposal, would appear to be of no use.

\begin{figure}
\epsfxsize=6cm
\hfil\epsfbox{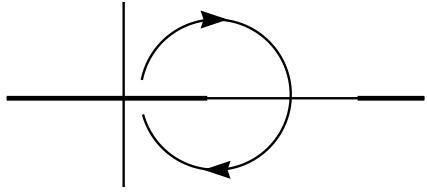}\hfill
\caption{The deformation of the integration contour into the complex $v\cdot\hat q$ plane.}
\label{fig:contour}
\end{figure}

The solution is to deform the contour away from the cut, into the complex $v\cdot\hat q$ plane, as shown in Fig.~\ref{fig:contour}.  Since the scale of momenta is set by $m_b$, the contour is now a distance $\sim m_b$ away from the resonances~\cite{CGG}.  Since $m_b\gg\lqcd$, it is reasonable to hope that a perturbative treatment in this region is valid.  Essentially, we are saved because we do not need to know $T^{\mu\nu}(q)$ for every value of $q$, just suitable integrals of $T^{\mu\nu}$.  That we can use such arguments to compute perturbatively the {\it average\/} value of a hadronic quantity, where at each point the quantity depends on nonperturbative physics, is known as (global) {\it parton-hadron duality.}  

Parton-hadron duality has the status of being somewhat more than an assumption, since it is known to hold in QCD in the limit $m_b\to\infty$, but somewhat less than an approximation, since it is not known how to compute systematically the leading corrections to it.  In any case, the limit $m_b\gg\lqcd$ plays a crucial role here.  By deforming the integration contour a distance $\sim m_b$ away from the resonance regime, we find the correspondence in QCD of our earlier intuitive stateprobabilityment: the probability of the decay products materializing as physical hadrons is unity, independent of the kinematics of the short distance process.  The local redistribution of probability in phase space due to the presence of hadronic resonances is irrelevant to the total decay.  Finally, we should note that since we do not have control over the corrections to local duality, it might work better in some processes than in others, for reasons that need not be apparent from within the calculation.  Hence one must be particularly wary of drawing dramatic conclusions from any surprising results of these inclusive calculations~\cite{FDW}.

\begin{figure}
\epsfxsize=8cm
\hfil\epsfbox{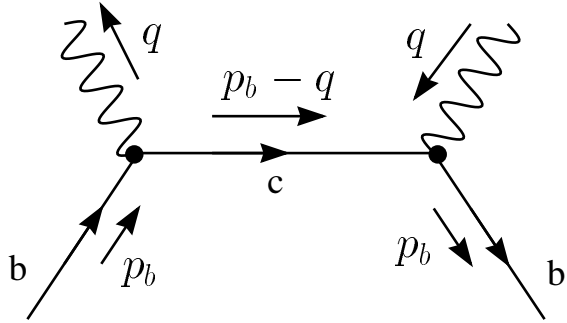}\hfill
\caption{The operator product expansion at tree level.}
\label{fig:top}
\end{figure}

Let us perform the operator product expansion at tree level, and for decay kinematics.  The Feynman diagram is given in Fig.~\ref{fig:top}, which yields the expression
\be
  T^{\mu\nu}=
  \bar b\gamma^\mu(1-\gamma^5)\,{\rlap/p_b-\rlap/q+m_c\over
  (p_b-q)^2-m_c^2+i\epsilon}\,\gamma^\nu(1-\gamma^5)b\,.
\ee
We now write
\bea
  p_b^\mu&=&m_bv^\mu +k^\mu=m_b(v^\mu+\hat k^\mu)\nonumber\\
  \hat q^\mu&=&q^\mu/m_b\nonumber\\
  \hat m_c&=&m_c/m_b\nonumber\\
  b(x)&=&e^{-im_bv\cdot x}\,h_v(x)+O(1/m_b)\,,
\eea
and expand in powers of $1/m_b$.  Since the operator product expansion is in terms of the effective field $h_v$, a factor of $k^\mu$ corresponds to an insertion of the covariant derivative $iD^\mu$.  Operator ordering ambiguities are to be resolved by considering graphs with external gluon fields.

As an example of the procedure, let us expand the propagator to order $1/m_b^2$.  (There are also corrections to the currents at this order, which are included in a full calculation.)  It is convenient to define the scaled hadronic invariant mass,
\be
  \hat s=(p_b-q)^2/m_b^2=1-2v\cdot\hat q+\hat q^2\,.
\ee
Then we find a contribution to $T^{\mu\nu}$ of the form
\be\label{Texpand}
  T^{\mu\nu}={1\over m_b}\,\bar h_v\gamma^\mu(\rlap/v-\rlap/{\hat q})
  \gamma^\nu(1-\gamma^5)\left[{1\over\hat s-\hat m_c^2+i\epsilon}
  +{2\hat k\cdot\hat q-\hat q^2\over
  (\hat s-\hat m_c^2+i\epsilon)^2}+\dots\;\right]h_v+\dots\,.
\ee
From this expression we can read off the operators which appear in the operator product expansion.  Since
\be
  {\rm Im}\;{1\over(\hat s-\hat m_c^2+i\epsilon)^n}\propto
  \delta^{(n-1)}(\hat s-\hat m_c^2)\,,
\ee
we see that the effect of taking the imaginary part in each term is to put the charm quark on its mass shell.  The leading term is a quark bilinear,
\be
  {1\over m_b}\,\bar h_v\gamma^\mu(\rlap/v-\rlap/{\hat q})
  \gamma^\nu(1-\gamma^5)h_v\,.
\ee
It is straightforward to compute its matrix element in the HQET using the trace formalism,
\be
  \langle B|\,\bar h_v\gamma^\mu(\rlap/v-\rlap/{\hat q})
  \gamma^\nu(1-\gamma^5)h_v\,|B\rangle
  =2m_B\left(2v^\mu v^\nu-g^{\mu\nu}-v^\mu\hat q^\nu-v^\nu\hat q^\mu
  +g^{\mu\nu}\,v\cdot\hat q +\dots\right)\,.
\ee
The ellipses denote terms of order $1/m_b^2$; there are no corrections of order $1/m_b$, by Luke's Theorem.  Finally, we contract the tensor $T^{\mu\nu}$ with $L_{\mu\nu}$ and perform the phase space integration (\ref{Tdef}).  In the end, the result is the same as we would have gotten directly by computing free quark decay.

Of course, if we only intended to reproduce the free quark decay result, we would never have introduced so much new formalism.  The value of the HQET framework is that it allows us to go beyond leading order and compute the next terms in the series in $1/m_b^n$.  For example, consider the operators induced by the expansion of the propagator (\ref{Texpand}).  The correction of order $1/m_b$ comes from the operator
\be
  {1\over m_b^2}\,{2\over(\hat s-\hat m_c^2+i\epsilon)^2}\,
  \bar h_v\gamma^\mu(\rlap/v-\rlap/{\hat q})
  \gamma^\nu(1-\gamma^5)\hat q\cdot iD\,h_v\,.
\ee
However, the matrix element of this operator is of the form
\be
  \langle B|\,\bar h_v\Gamma^\alpha (v,q)\,iD_\alpha\,h_v\,|B\rangle\,,
\ee
which, as we have seen, vanishes by the classical equation of motion.  In fact, since all $1/m_b$ corrections, from any source, have a single covariant derivative, they all vanish in the same way.  This is the analogue of Luke's Theorem for inclusive decays~\cite{CGG}.  The correction of order $1/m_b^2$ in Eq.~(\ref{Texpand}) is
\be
  -{1\over m_b^3}\,{1\over(\hat s-\hat m_c^2+i\epsilon)^2}\,
  \bar h_v\gamma^\mu(\rlap/v-\rlap/{\hat q})
  \gamma^\nu(1-\gamma^5)(iD)^2\,h_v\,.
\ee
The matrix element of this operator is related by the heavy quark symmetry to $\lambda_1$, the expectation value of ${\cal O}_K$.  The full expansion of $T^{\mu\nu}$ also induces operators with explicit factors of the gluon field, whose matrix elements are related to $\lambda_2$.

We now present the result for the inclusive semileptonic decay rate, up to order $1/m_b^2$ in the heavy quark expansion, and with the complete radiative correction of order $\alpha_s$.  We also include that part of the two loop correction which is proportional to $\beta_0\alpha_s^2$.  Since $\beta_0\approx9$, perhaps this term dominates the two loop result.  In any case, it is interesting for other reasons, as we will see below.

Let us first consider the decay $B\to X_u\,\ell\bar\nu$, for which the decay rate simplifies since $m_u=0$.  We find~\cite{SV,MW,GPR,LSW1,FLS96}
\bea\label{burate}
  \Gamma(B\to X_u\,\ell\bar\nu)={G_F^2|V_{ub}|^2\over192\pi^3}\,
  m_b^5\,\Bigg[1&+&\left({25\over6}-{2\pi^2\over3}\right)
  {\alpha_s(m_b)\over\pi}
  -(2.98\beta_0+C_u)\left({\alpha_s(m_b)\over\pi}\right)^2+\dots\nonumber\\
  &+&{\lambda_1-9\lambda_2\over2m_b^2}+\dots\;\Bigg]\,.
\eea
When we include the charm mass, it is convenient to write the unknown quark masses in terms of the measured meson masses and the parameters of the HQET.  In terms of the spin averaged mass $\overline m_B=(m_B+3m_{B^*})/4$, we have
\be\label{mexpand}
  m_b=\overline m_B-\bar\Lambda+{\lambda_1\over2m_B}+\dots\,,
\ee
and analogously for $m_c$.  We then find~\cite{SV,MW,Nir,LSW2}
\bea\label{bcrate}
  \Gamma(B\to X_c\,\ell\bar\nu)=&&{G_F^2|V_{cb}|^2\over192\pi^3}
  \,m_B^5\times 0.369\,
  \Bigg[1-1.54{\alpha_s(m_b)\over\pi}
  -(1.43\beta_0+C_c)\left({\alpha_s(m_b)\over\pi}\right)^2+\dots\nonumber\\
  &&\mbox{}-1.65{\bar\Lambda\over m_B}
  \left(1-0.87{\alpha_s(m_b)\over\pi}\right)
  -0.95{\bar\lambda^2\over m_B^2}-3.18{\lambda_1\over m_B^2}
  +0.02{\lambda_2\over m_B^2}+\dots\;\Bigg]\,.\nonumber\\
\eea
All the coefficients which appear in this expression are known functions of $\overline m_D/\overline m_B$, and are evaluated at the physical point $\overline m_D/\overline m_B=0.372$.  In both $B\to X_u\,\ell\bar\nu$ and $B\to X_c\,\ell\bar\nu$, the power corrections proportional to $\lambda_1$ and $\lambda_2$ are numerically small, at the level of a few percent.

\subsection{Renormalons and the pole mass}

The inclusive decay rate depends on the heavy quark mass $m_b$, either explicitly, as in Eq.~(\ref{burate}), or implicitly through $\bar\Lambda$, as in Eq.~(\ref{bcrate}).  At tree level, $m_b$ is just the coefficient of the $\bar b\,b$ term in the QCD Lagrangian, but beyond that we are faced with the question of what exactly we mean by $m_b$.  Should we take an $\overline{\rm MS}$ mass, such as $\overline m_b(m_b)$?  Or should we take the pole mass $\mbpole$, or maybe some other quantity?  The various prescriptions for $m_b$ can vary by hundreds of MeV, and, since the total rate is proportional to $m_b^5$, the question is of practical importance if we hope to make accurate phenomenological predictions.

At a fixed order in QCD perturbation theory, the answer is clear.  The heavy quark masses which appear come from poles in quark propagators, so we should take $\mbpole$ (and $\mcpole$).  This is also the prescription for the mass which cancels out the on-shell part of the heavy quark field in the construction of $\lhqet$.  Hence the difference of heavy quark pole masses is known quite well,
\bea
  \mbpole-\mcpole&=&\left(\overline m_B-\bar\Lambda+{\lambda_1\over2m_B}+\dots
  \right)-\left(\overline m_D-\bar\Lambda+{\lambda_1\over2m_D}+\dots\right)
  \nonumber\\
  &=&3.34\gev+O(\lqcd^2/m_Q^2)\,.
\eea
Since $\Gamma(B\to X_c\,\ell\bar\nu)$ depends approximately as $m_b^2(m_b-m_c)^3$, the uncertainty due to quark mass dependence is reduced.

The problem, of course, is that there is no sensible nonperturbative definition of $\mbpole$, since due to confinement there is no actual pole in the quark propagator.  Hence a direct experimental determination of a value for $\mbpole$ to insert into the theoretical expressions (\ref{burate}) and (\ref{bcrate}) is not possible.  How, then, can we do phenomenology?

\begin{figure}
\epsfxsize=12cm
\hfil\epsfbox{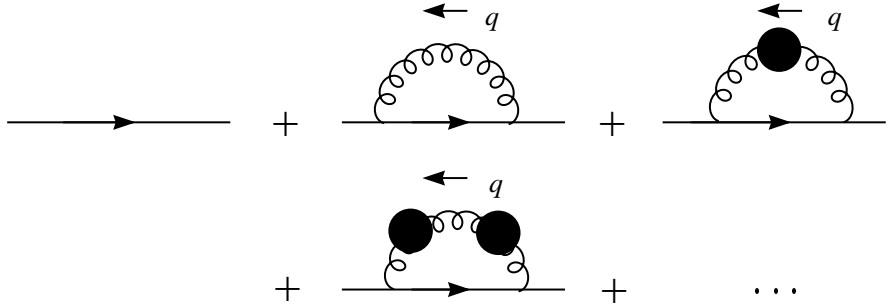}\hfill
\caption{The radiative corrections to $\mbpole$ of order $\alpha_s(m_b)^{n+1}\beta_0^n$.}
\label{fig:mbubbles}
\end{figure}

One approach would be to define $\mbpole$ to be the pole mass as computed in perturbation theory, truncate at some order, and then estimate the theoretical error from the uncomputed higher order terms.  However, it turns out that even {\it within\/} perturbation theory the concept of a quark pole mass is ambiguous.  Consider a particular class of diagrams which contribute to $\mbpole$, shown in Fig.~\ref{fig:mbubbles}.  The perturbation theory is developed as an expansion in the small parameter $\alpha_s(m_b)$, so we hope that it will be well behaved.  Each of the bubbles represents an insertion of the gluon self-energy, which is proportional at lowest order to $\alpha_s(m_b)\beta_0$.  Of course, the infinite sum of the graphs in Fig.~\ref{fig:mbubbles} can be absorbed into the one loop graph, with a compensating change in the coupling from $\alpha_s(m_b)$ to $\alpha_s(q)$, where $q$ is the loop momentum.
The result is an expansion for $\mbpole$ of the form
\be
  \mbpole=\overline m_b(m_b)\bigg[1+a_1\alpha_s(m_b)+
  (a_2\beta_0+b_2)\alpha_s^2(m_b)
  +(a_3\beta_0^2+b_3\beta_0+c_3)\alpha_x^3(m_b)+\dots\bigg]\,.
\ee
The graphs in Fig.~\ref{fig:mbubbles} contribute the terms proportional to $\alpha^{n+1}_s(m_b)\beta_0^n$.  Since $\beta_0\approx9$ these terms are ``intrinsically'' larger than ones with fewer powers of $\beta_0$, and we might hope that their sum approximates the full series.  However, it is important to realize that the only limit of QCD in which such terms actually dominate is that of large number of light quark flavors, in which case the sign of $\beta_0$ is opposite to that of QCD.  Although this is a physical limit of an abelian theory, we are certainly not close to that limit here.  The ansatz of keeping only the terms proportional to $\alpha^{n+1}_s(m_b)\beta_0^n$ is known as ``naive nonabelianization'' (NNA)~\cite{BBZ}.

What is most interesting about the series of terms shown in Fig.~\ref{fig:mbubbles}, $\sum a_n\alpha^n_s(m_b)\beta_0^{n-1}$, is that it does not converge.  Already in the graphs kept in the NNA ansatz, we are sensitive to the fact that QCD is an asymptotic, rather than a convergent expansion.  For large $n$ the coefficients $a_n$ diverge as $n!$, much stronger than any convergence due to the powers $\alpha^n_s(m_b)$.  The series can only be made meaningful if this divergence is subtracted.  As with many subtraction prescriptions, there is a residual finite ambiguity.\footnote{In a  formal treatment, this ambiguity arises from a choice of contour in the Borel plane.}  This ambiguity, known as an ``infrared renormalon'', leads to an ambiguity in the pole mass of order~\cite{BBZ,Bigi,NS}
\be
  \delta\mbpole\sim100\mev\,.
\ee
By the definition (\ref{lambardef}), $\bar\Lambda$ also inherits this ambiguity.

The expressions (\ref{burate}) and (\ref{bcrate}) are plagued by two problems.  The first is the renormalon ambiguity in $\mbpole$ and $\bar\Lambda$.  The second is that the perturbative expansion for the rate $\Gamma$ is itself divergent, and {\it also\/} has an infrared renormalon.  In the expansion
\be
  \Gamma=\Gamma_0\left[\sum a'_n\alpha^n_s(m_b)\beta_0^{n-1}+
  {\rm (power\ corrections)\ }\right]\,,
\ee
the coefficients $a_n'$ also diverge as $n!$.  However, it turns out that these two problems actually cure each other, because the infrared renormalons in $\mbpole$ and in the perturbation series for $\Gamma$ cancel~\cite{Bigi,LMS}.  We can exploit this cancelation to improve the predictive power of the theoretical computation of the rate.  Without this improvement, the infrared renormalons render the expressions (\ref{burate}) and (\ref{bcrate}) of dubious phenomenological utility.

The most reliable approach, theoretically, is to eliminate $\mbpole$ or $\bar\Lambda$ explicitly from the rate by computing and measuring another quantity which also depends on it.  For example, let us consider the charmless decay rate $\Gamma(B\to X_u\,\ell\bar\nu)$ and the average invariant mass $\langle s_H\rangle$ of the hadrons produced in the decay.  Each of these expressions suffers from a poorly behaved perturbation series in the NNA approximation.  Ignoring terms of relative order $1/m_b^2$ and writing the rate in terms of $\bar\Lambda$ instead of $\mbpole$, we find to five loop order~\cite{BBZ},
\bea
  \Gamma&=&{G_F^2|V_{ub}|^2\over192\pi^3}\,m_B^5
  \Bigg[1-2.41{\alpha_s\over\pi}
  -2.98\left({\alpha_s\over\pi}\right)^2\beta_0
  -4.43\left({\alpha_s\over\pi}\right)^3\beta_0^2\nonumber\\
  &&\qquad\qquad\qquad\mbox{}
  -7.67\left({\alpha_s\over\pi}\right)^4\beta_0^3
  -15.7\left({\alpha_s\over\pi}\right)^5\beta_0^4+\dots
  -5{\bar\Lambda\over m_B}+\dots\Bigg]\nonumber\\
  &=&{G_F^2|V_{ub}|^2\over192\pi^3}\,m_B^5\,\Big[
  1-0.061-0.120-0.107-0.111-0.136+\dots\nonumber\\
  &&\qquad\qquad\qquad\mbox{}-5\bar\Lambda/m_B+\dots\Big]\,,
\eea
for $\alpha_s(m_b)=0.21$ and $\beta_0=9$.  As we see, not only does the perturbation series fail to converge, it does not even have an apparent smallest term, where one should truncate to minimize the error of the asymptotic series.  The series for $\langle s_H\rangle$ exhibits a similar behavior~\cite{FLS96},
\bea
  \langle s_H\rangle&=&m_B^2\Bigg[0.20{\alpha_s\over\pi}
  +0.35\left({\alpha_s\over\pi}\right)^2\beta_0
  +0.64\left({\alpha_s\over\pi}\right)^3\beta_0^2
  +1.29\left({\alpha_s\over\pi}\right)^4\beta_0^3\nonumber\\
  &&\qquad\qquad\qquad\mbox{}
  +2.95\left({\alpha_s\over\pi}\right)^5\beta_0^4+\dots
  +{7\over10}{\bar\Lambda\over m_B}+\dots\Bigg]\nonumber\\
  &=&m_B^2\,\left[
  0.0135+0.0141+0.0156+0.0189+0.0261+\dots-7\bar\Lambda/10m_B+\dots\right]\,.
\eea
However, the situation improves dramatically if we eliminate $\bar\Lambda$ and write $\Gamma$ directly in terms of $\langle s_H\rangle$,
\be
  \Gamma={G_F^2|V_{ub}|^2\over192\pi^3}\,m_B^5\left[
  1-7.14{\langle s_H\rangle\over m_B^2}-0.064-0.020-0.0002-0.022-0.047+\dots
  \right]\,.
\ee
By truncating this series at its smallest term, 0.0002, we obtain a new expression in which the theoretical errors are under control.  The price is that we must now measure a second quantity, $\langle s_H\rangle$, in the same decay.

An analogous approach works for inclusive decays to charm, in which case the new quantity which we must measure is $\langle s_H-\overline m_D^2\rangle$.  Let us apply this analysis to the extraction of $V_{cb}$ from the inclusive decay $B\to X_c\,\ell\bar\nu$.  Eliminating $\bar\Lambda$ as before, we find~\cite{FLS96}
\bea\label{width2}
  \Gamma(B\to X_c\ell\nu)&=&{G_F^2|V_{cb}|^2\over192\pi^3} m_B^5\times0.369
  \bigg[1-1.17{\alpha_s(m_b)\over\pi}-0.74\beta_0
  {\alpha_s^2(m_b)\over\pi^2}\nonumber\\
  &&\qquad\qquad\qquad\qquad\qquad\mbox{}
  -7.17{\langle s_H-\overline m_D^2\rangle
  \over m_B^2}+\dots\bigg]\,,
\eea
plus small terms of order $\lambda_{1,2}/m_B^2$.  The current data on $\langle s_H-\overline m_D^2\rangle$ are still inconclusive, but they will improve in the future.  As an illustration of the utility of the method, however, let us consider the physical limit $\langle s_H-\overline m_D^2\rangle\ge0$.  In that case, we already can obtain the bound
\be
  V_{cb}\ge0.037\times\left({\tau_B\over1.54\,{\rm ps}}\right)^{-1/2}\,,
\ee
where we have used the observed $B$ semileptonic branching ratio of $10.7\%$.
As the data on $\langle s_H-\overline m_D^2\rangle$ improve, so will this bound.  Note that it is entirely consistent with the value (\ref{vcbexcl}) obtained from exclusive decays.

An alternative approach is to express the width $\Gamma$ in terms of the running mass $\overline m_b(m_b)$ instead of another inclusive observable~\cite{Bigi,BBB}.  Since the $\overline{\rm MS}$ mass is a short distance quantity, this also eliminates the infrared renormalon, which is associated with long distance physics.  However, from a phenomenological point of view, it begs the question of how the running mass is to be determined from experiment.  Possibilities include quarkonium spectroscopy, QCD sum rules, and lattice calculations, but in all of these cases it is unclear both how to determine reliably the accuracy of the method, and how to deal with renormalon ambiguities in a manner that is {\it consistent\/} with their treatment in the calculation of $\Gamma$.  With additional progress, however, such an approach might eventually prove fruitful.

\section{Heavy Hadron Production via Fragmentation}

So far we have applied the idea of heavy quark symmetry to heavy hadron spectroscopy and to heavy hadron decay.  We can also apply it to heavy hadron production via fragmentation.  Although there is not yet available the wealth of detailed data for production processes that there is for spectroscopy and decays, the study of heavy quark fragmentation is an elegant application of the heavy quark limit in a new regime~\cite{FP}.

\subsection{The physics of heavy quark fragmentation}

The production of a heavy hadron proceeds in two steps.   First, the
heavy quark itself must be created; because of its large mass, this
process takes place over a time scale $\tau_Q$ which is very short, $\tau_Q\le1/m_Q$.  Second, some light degrees of freedom assemble themselves about the heavy quark to make a color neutral heavy hadron, a process which involves nonperturbative strong interactions and typically takes much
longer, $\tau_F\sim1/\lqcd\gg\tau_Q$.  If the heavy quark is produced with a large velocity in the center of mass frame, and if there is plenty of available energy, then production of these light degrees of freedom will be local in phase space and independent of the initial state.  This is the fragmentation regime.  We will see that heavy quark symmetry simplifies the description of heavy hadron production via fragmentation, because, as before, it allows us to separate certain properties of the heavy quark from those of
the light degrees of freedom.  This is particularly important in the
production of excited heavy hadrons, for which the behavior of the
spin of the light degrees of freedom can be quite interesting.

Heavy quark symmetry simplifies the picture of fragmentation in several ways.  First, since the fragmentation process involves the exchange only of soft momenta, the velocity $v^\mu$ of the hadron $H_Q$ which is produced is the same as the velocity of the heavy quark $Q$.  Second, the angular momenta $S_Q$ and $J_\ell$ of the heavy quark and the light degrees of freedom are well defined and can be followed individually through the processes of production and decay.  Third, the typical precession time $\tau_S$ of $S_Q$ and $J_\ell$ about each other is of order $\tau_S\sim m_Q/\lqcd^2\gg1/\lqcd$, since the precession is induced by the chromomagnetic interaction.  Hence $\tau_S\gg\tau_F$, and during the fragmentation process $S_Q$ can essentially be taken to be frozen in direction.

Suppose that $Q$ fragments initially to a state with light degrees of freedom with spin $J_\ell$.  This could correspond to two possible hadronic states: $H_Q$, with spin $J=J_\ell-{1\over2}$, or $H_Q^*$, with spin $J=J_\ell+{1\over2}$.  Let the states $H_Q^{(*)}$ have lifetime $\tau$.  How should we treat the interaction of $S_Q$ with $J_\ell$?

The answer depends critically on the relationship of $\tau$ to the precession time $\tau_S$.  While the order of magnitude of $\tau_S$ is essentially fixed at $m_Q/\lqcd^2$, the lifetime $\tau$ can vary widely, depending on the channel in which $H_Q^{(*)}$ primarily decays.  Of course, if $H_Q^{(*)}$ decays weakly, then $\tau\gg\tau_S$.  But if $H_Q^{(*)}$ decays strongly, then $\tau$ depends critically on the phase space available for the transition.  In the absence of phase space factors, $\tau\sim1/\lqcd\ll\tau_S$.  But since the pion, the lightest hadron, is not massless, it is possible to have a situation where an allowed strong decay is either forbidden by phase space (as with $B^*\to B+\pi$), or severely suppressed (as with $D^*\to D+\pi$).

The treatment of a given hadron doublet depends on the relative size of $\tau$ and $\tau_S$.  Let us consider the two extreme possibilities.  The first corresponds to a strong decay with plenty of phase space, so $\tau_S\gg\tau$.  Here $H_Q$ and $H_Q^*$ are formed and then decay before the angular momenta $S_Q$ and $J_\ell$ have a chance to interact.  The fact that there is no precession means that there is no change in the orientation of $S_Q$ and $J_\ell$.  If, for example, either spin were polarized before or during the fragmentation process, this polarization would be undisturbed by passing through the resonance $H_Q^{(*)}$.  The strong decay of the heavy hadron then would reflect the orientation of the light degrees as they were produced.

Note that it is the very same spin exchange interaction which is inhibited here which is responsible for the splitting $\Delta_H$ between $H_Q$ and $H_Q^*$.  Hence, under these conditions the resonances are almost completely {\it overlapping,} with widths $\Gamma=1/\tau$ satisfying $\Gamma\gg\Delta_H$.  This is another consequence of the effective decoupling of $S_Q$ and $J_\ell$, which are independent good quantum numbers of the resonances.  It also provides a convenient criterion for determining directly whether we are in the regime $\tau_S\gg\tau$: we cannot distinguish separate resonances for $H_Q$ and $H_Q^*$.

The second possibility is the opposite extreme, $\tau_s\ll\tau$.
This corresponds to heavy hadrons which decay weakly or
electromagnetically, or to strong decays which are severely
suppressed by phase space.  Here the spins $S_Q$ and $J_\ell$ have
plenty of time to interact, precessing about each other many times
before $H_Q$ and $H_Q^*$ decay.  There is at least a partial degradation
of any initial polarization of $S_Q$, as well as a degradation of any
information about the fragmentation process which may be carried by
the light degrees of freedom.  The signature of this situation is
that the states $H_Q$ and $H_Q^*$ are well separated resonances, since
the chromomagnetic interactions have ample opportunity to produce a
hyperfine splitting much larger than the width,
$\Delta_H\gg\Gamma$.  In contrast with the first case, here the
heavy and light spins are resolved into states of definite total
spin $J=|J_\ell\pm {1\over2}|$.  This resolution destroys information about the individual angular momenta $S_Q$ and $J_\ell$, and the decays of $H_Q$ and $H_Q^*$ only partially reflect the features of the production process.

\subsection{Production and decay of $D_1$ and $D_2^*$}

We will illustrate these ideas with two examples.  The first concerns the production and decay of the excited charmed mesons $D_1$ and $D_2^*$.  These are states with light degrees of freedom with angular momentum and parity $J_\ell^P={3\over2}^+$; in the quark model, this would correspond to a light antiquark with a unit of orbital angular momentum.  The physical states are the $D_1$, with mass $2423\mev$ and width $18\mev$, and the $D_2^*$, with mass $2458\mev$ and width $21\mev$.  The splitting between them is $35\mev$; although this is not much larger than the individual widths, for simplicity we will treat them in this limiting case $\tau\gg\tau_S$.

Let us review the sequence of events.  First, the charm quark is created in some hard interaction, with a time scale $\tau_c\sim1/m_c$.  Second, light degrees of freedom with $J_\ell={3\over2}$ are created in a fragmentation process, over a time $\tau_F\sim1/\lqcd$, forming a color neutral charmed hadron.  Third, the angular momenta $S_c$ and $J_\ell$ precess about each other, resolving the individual $D_1$ and $D_2^*$ states, which requires a time $\tau_S\sim m_c/\lqcd$.  Then, finally, after a time $\tau$ greater than $m_c/\lqcd$, the $D_1$ or $D_2^*$ emits a pion and decays to a $D$ or $D^*$.
One question we might ask about this process is, what is the alignment of the light degrees of freedom produced in the fragmentation?  Of course, the answer depends on nonperturbative QCD and is difficult to calculate from first principles.  Still, we can explore the question experimentally. 

Since the total angular momentum of the light degrees of freedom is $J_\ell={3\over2}$, they can be produced with
helicity $h=\pm{3\over2}$ or $h=\pm{1\over2}$ along the fragmentation
axis.  While parity invariance of the strong interactions requires
that the probabilities for helicities $h$ and $-h$ are identical,
the relative production of light degrees of freedom with
$|h|={3\over2}$ versus $|h|={1\over2}$ is determined in some
complicated way by nonperturbative dynamics.  Let the
quantity $w_{3/2}$ denote the probability that $|h|=\case32$,
\be
  w_{3/2}=P(h=\textstyle{3\over2})+P(h=-\textstyle{3\over2})\,.
\ee
Then $1-w_{3/2}$ is the probability that $|h|={1\over2}$.  Completely
isotropic production corresponds to $w_{3/2}={1\over2}$.  The observable $w_{3/2}$ is a new nonperturbative parameter of QCD, which is well defined only in the heavy quark limit.

This new parameter can be measured directly in the strong decay of the
$D_2^*$ or $D_1$.  For example, consider the angular distribution
of the pion with respect to the fragmentation axis in the decay
$D_2^*\to D+\pi$.  This is a decay of the light degrees of freedom
in the excited hadron, so it will depend on their initial
orientation (that is, on $w_{3/2}$) and on the details of the
precession of $J_\ell$ around $S_Q$ during the lifetime of the
$D_2^*$.  Following the direction of $J_\ell$ through the sequence of
fragmentation, precession and decay, we find the distribution~\cite{FP}
\be\label{d2todpi}
  {1\over\Gamma}{{\rm d}\Gamma\over{\rm d}\cos\theta}=
  \case14\left[1+3\cos^2\theta-6w_{3/2}
  (\cos^2\theta-\case13)\right]\,.
\ee
This distribution is isotropic only when $w_{3/2}=\case12$, that
is, when the light degrees of freedom are produced isotropically in
the fragmentation process.  Similar distributions are found in the
decays $D_2^*\to D^*+\pi$ and $D_1\to D^*+\pi$.

A fit of ARGUS data~\cite{AR2} to the expression (\ref{d2todpi}) seems
to indicate that a small value of $w_{3/2}$ is preferred; while the
errors are large, we find that
$w_{3/2}<0.24$ at the 90\% confidence level.  It would be
nice to confirm this result with a sharper measurement, and not only for
the charmed mesons but in the bottom system as well.  Since $w_{3/2}$
is intrinsically nonperturbative, we do not have any real
theoretical understanding of why it should be small, although
one can construct a qualitative understanding based on a string picture of fragmentation~\cite{Chow}.

\subsection{Polarization of $\Lambda_b$ at SLC/LEP}

The second example springs from a practical question:  What is the polarization of $\Lambda_b$ baryons produced at the $Z$ pole?  This question is motivated by the fact that $b$ quarks produced in the decay of the
$Z$ are 94\% polarized left-handed.  Since the $\Lambda_b$ is composed of a $b$ quark and light degrees of freedom with $J_\ell=0$, the orientation of a $\Lambda_b$ is identical to the orientation of the $b$ quark inside it.  Similarly, the $b$ quark spin does not precess inside a $\Lambda_b$.  Hence if a $b$ quark produced at the $Z$ fragments to a $\Lambda_b$, then those baryons should inherit the left-handed polarization of the
quarks and reveal it in their weak decay.

Unfortunately, life is not that simple.  Two recent measurements of
$\Lambda_b$ polarization $P(\Lambda_b)$ from LEP are~\cite{DE2,AL2}
\bea
  {\rm DELPHI:}&&\qquad 0.08\,^{+0.35}_{-0.29}{\rm (stat.)}\,^{+0.18}_{-0.16}
  {\rm (syst.)}\nonumber\\
  {\rm ALEPH:}&&\qquad 0.26\,^{+0.20}_{-0.25}{\rm (stat.)}\,^{+0.12}_{-0.13}
  {\rm (syst.)}\,,\nonumber
\eea
both a long way from $P(\Lambda_b) =0.94$.  The reason is that not all $b$ quarks which wind up as $\Lambda_b$ baryons get there directly.  There is a competing process, in which they fragment first to the excited baryons $\Sigma_b$ and $\Sigma_b^*$, which then decay to $\Lambda_b$ via pion emission.  These new states have light degrees of freedom with $J_\ell=1$.  If they have a lifetime $\tau>\tau_S$, then the $b$ quark will have time to precess about $J_\ell$ and its polarization will be degraded.  The result will be a net sample of $\Lambda_b$'s with a polarization less than 94\%, as is in fact observed.

In addition to the requirement that $\tau>\tau_S$ for the $\Sigma_b^{(*)}$, any depolarization of $\Lambda_b$'s by this mechanism depends on two unknown quantities.  First, there is the production rate $f$ of $\Sigma_b^{(*)}$ relative to $\Lambda_b$.  Isospin and spin counting enhance $f$ by a factor
of nine, while the mass splitting between $\Sigma_b^{(*)}$ and $\Lambda_b$ suppresses it; studies based on the Lund Monte Carlo ~\cite{Lund} indicate $f\approx0.5$ with a very large uncertainty.  Of course, it would be preferable to measure $f$ directly. Second, there is the orientation of the spin $J_\ell$ with respect to the fragmentation axis.  This orientation, which is nonperturbative in origin, allows the helicities $h=1,0,-1$.  By analogy with the treatment of the $D_1$ and $D_2^*$, let us define
\be
  w_1=P(h=1)+P(h=-1)\,.
\ee
In this case, isotropic production corresponds to $w_1={2\over3}$.  We may measure $w_1$ from the angle of the pion with respect to the fragmentation
axis in the decay $\Sigma_b^*\to\Lambda_b+\pi$~\cite{FP},
\be\label{sigtolampi}
  {1\over\Gamma}{{\rm d}\Gamma\over{\rm d}\cos\theta}=
  \case14\left[1+3\cos^2\theta-\case92w_1
  (\cos^2\theta-\case13)\right]\,.
\ee
It turns out that the decay $\Sigma_b\to\Lambda_b+\pi$ is isotropic in
$\cos\theta$ for any value of $w_1$.

The polarization retention of the $\Lambda_b$ may be computed in terms of $f$ and $w_1$.  As before, we will consider the simpler situation in which the $\Sigma_b$ and the $\Sigma_b^*$ do not overlap, so $\tau\gg\tau_S$.  Then the polarization of the observed $\Lambda_b$'s is $P(\Lambda_b)=R(f,w_1)P(b)$, where $P(b)=94\%$ is the initial polarization of the $b$ quarks, and~\cite{FP}
\be
  R(f,w_1) = {1+\case19(1+4w_1)f\over1+f}\,.
\ee
Note that for $f=0$ (no $\Sigma_b^{(*)}$'s are produced), $R(0,w_1)=1$ and there is no depolarization.  For the Lund value $f=0.5$, $R$ ranges between 0.70 and 0.85 for $0\le w_1\le 1$.

Can the very low measured values of $P(\Lambda_b)$ be accommodated by the present data on the $\Sigma_b^{(*)}$?  The situation is still unclear, because not much is known about the $\Sigma_b$ and $\Sigma_b^*$ states.  DELPHI~\cite{DE3} has reported the observation of these states, but with masses which deviate significantly from those that one would expect based on heavy quark symmetry and the observed $\Sigma_c$ and $\Sigma_c^*$~\cite{Falk96}.  Along with the masses, DELPHI also reports $w_1\approx0$ and $1<f<2$.  If this is confirmed, then a polarization in the range $P(\Lambda_b)\approx40\%-50\%$ is easy to accommodate.  On the other hand, the CLEO analysis~\cite{CL2} of the decay of the $\Sigma_c^*$ yields $w_1=0.71\pm0.13$, consistent with isotropic fragmentation.  Recall that by heavy quark symmetry, $w_1$ measured in the charm and bottom systems must be the same, so this result is inconsistent with the report from DELPHI.  Clearly, further measurements are needed to resolve this situation.

\section{Concluding Remarks}

Unfortunately, we have had time in these lectures only to introduce a very few of the many applications of heavy quark symmetry and the HQET to the physics of heavy hadrons.  Since its development less than ten years ago, it has become one of the basic tools of QCD phenomenology.  Much of the popularity and utility of the HQET certainly come from its essential simplicity.  The elementary observation that the physics of heavy hadrons can be divided into interactions characterized by short and long distances gives us immediately a clear and compelling intuition for the properties of heavy-light systems.  The straightforward manipulations which lead to the HQET then allow this intuition to form the basis for a new systematic expansion of QCD.  The deeper understanding of heavy hadrons which we thereby obtain will become increasingly important as the end of the second millennium approaches and the $B$ Factory Era begins.

\section*{Acknowledgements}
I would like to thank the organizers of the Summer Institute for the opportunity to present these lectures, and for arranging a most interesting and pleasant summer school.  This work was supported by the National Science Foundation under Grant No.~PHY-9404057 and National Young Investigator Award No.~PHY-9457916, by the Department of Energy under Outstanding Junior Investigator Award No.~DE-FG02-94ER40869, and by the Alfred P.~Sloan Foundation.

\end{document}